\documentclass[twocolumn]{aastex6}
\slugcomment{To Appear in The Astrophysical Journal}

\shorttitle{New Neutron-Capture Measurements in 23 Open Clusters. I. The R-Process}
\shortauthors{Overbeek et al.}

\begin{document}

\title{New Neutron-Capture Measurements in 23 Open Clusters. I. The R-Process}
\author{Jamie C. Overbeek}
\affil{Indiana University Astronomy Department, Swain West 318, 727 East 3rd Street, Bloomington, IN 47405, USA; joverbee@indiana.edu}

\author{Eileen D. Friel}
\affil{Indiana University Astronomy Department, Swain West 318, 727 East 3rd Street, Bloomington, IN 47405, USA}

\author{Heather R. Jacobson}
\affil{Massachusetts Institute of Technology, Kavli Institute for Astrophysics and Space Research, Cambridge, MA 02139, USA}

\begin{abstract}

Neutron-capture elements, those with Z $>$ 35, are the least well-understood in terms of nucleosynthesis and formation environments. The rapid neutron-capture, or r-process, elements are formed in the environments and/or remnants of massive stars, while the slow neutron-capture, or s-process, elements are primarily formed in low-mass AGB stars. These elements can provide much information about Galactic star formation and enrichment, but observational data is limited. We have assembled a sample of 68 stars in 23 open clusters that we use to probe abundance trends for six neutron-capture elements (Eu, Gd, Dy, Mo, Pr, and Nd) with cluster age and location in the disk of the Galaxy. In order to keep our analysis as homogenous as possible, we use an automated synthesis fitting program, which also enables us to measure multiple (3-10) lines for each element. We find that the pure r-process elements (Eu, Gd, and Dy) have positive trends with increasing cluster age, while the mixed r- and s- process elements (Mo, Pr, and Nd) have insignificant trends consistent with zero. Pr, Nd, Eu, Gd, and Dy have similar, slight (though mostly statistically significant) gradients of $\sim$0.04 dex/kpc. The mixed elements also appear to have nonlinear relationships with R$_{\mathrm{GC}}$.

\end{abstract}


\section{Introduction}

\indent Various astronomical fields such as stellar evolution, the formation of the Galaxy, and galactic chemical evolution all require an understanding of stellar nucleosynthesis. Although the light elements, $\alpha$- and Fe-peak element nucleosynthesis processes are relatively well understood, less theoretical and observational data are available for elements heavier than Z $\sim$ 35 which are collectively called the neutron-capture elements. These elements build in two main ways: via slow neutron capture (the s-process) and rapid neutron-capture (the r-process), with most having contributions from both processes. Once an atomic nucleus captures a neutron, it will then undergo $\beta$-decay until it reaches the valley of $\beta$-stability. The slow neutron-capture process occurs in environments with relatively low neutron flux where $\beta$-decay occurs more quickly than successive neutron captures. The r-process occurs in high neutron flux environments where nuclei capture neutrons much more quickly than the timescale of $\beta$-decay. Thus, the two processes create different isotopes with varying efficiency. 
\\
\indent The s-process is comprised of several components, the weak, the strong, and the main, with recent evidence for a light element primary process \citep{Travaglio04}. s-process elements are produced in multiple astrophysical sites, although low-mass (M $< 4$M$_{\odot}$) AGB stars are generally thought to be the most significant contributors. The site of r-process formation, however, has not yet been conclusively identified; core-collapse supernovae and neutron star-neutron star or neutron star-black hole mergers are the primary candidates \citep[see, e.g.,][]{Burbidge57, Freiburghaus99, Vangioni16}. It has also been suggested that there may be two r-process sites, one for species below and one for species above A $\sim$ 140 \citep{Wasserburg96}.
\\
\indent Open clusters (OCs) are useful tools for studying neutron-capture abundances in approximately solar-metallicity stars. Their ages and locations within the Galaxy are relatively easy to determine and cover a significant range in parameter space. Though much work has been done measuring r-process element abundances in metal-poor halo stars \citep[e.g.][]{Spite78, Sneden94, Honda04, Francois07}, relatively few measurements of OC r-process abundances are present in the literature, mostly focusing on Eu. Unfortunately, these Eu results are sometimes conflicting. \citet{Yong12}, using a sample of ten OCs plus four literature measurements, find a [Eu/Fe] trend with age of -0.01 $\pm$ 0.01 dex Gyr$^{-1}$, while \citet{JF13}, using a sample of 19 OCs, find a trend in the opposite sense of +0.023 dex Gyr$^{-1}$.
\\
\indent In order to better constrain the high-metallicity r-process regime, we have assembled a sample of 68 stars in 23 OCs and measured abundances for six neutron-capture elements (three primarily r-process and three mixed r- and s-process elements) which we present here. S-process abundance measurements for this sample will be discussed in a forthcoming paper. The article is laid out as follows: in Section \ref{sec:data} we discuss our sample data, in Section \ref{sec:params} we describe atmospheric parameter determinations, in Section \ref{sec:abuns} we give the neutron-capture abundance measurements and dispersions, in Section \ref{sec:discr} we discuss anomalous stars, Section \ref{sec:errors} describes our error estimates, in Section \ref{sec:discussion} we draw comparisons to the literature and discuss trends and gradients, and in Section \ref{sec:summary} we summarize our findings.

\section{Observations and Data}  \label{sec:data}
\subsection{Previously Presented and Archival Data}  \label{sec:prevdata}

\indent Our OC data set is comprised mainly of KPNO 4m, CTIO, and HET spectra taken in past observing runs \citep[described in][]{Friel03, Friel05, Jacobson08, Jacobson09, Friel10, JF13}. These are observations of red giants in Be 17, Be 39, Cr 261, M67, NGC 188, NGC 1193, NGC 1245, NGC 1817, NGC 1883, NGC 2141, NGC 2158, NGC 2194, NGC 2355, NGC 6939, and NGC 7142.
\\
\indent We have also obtained Keck spectra of five clusters, Be 18, Be 21, Be 22, Be 32, and PWM4, and CTIO spectra for Be 31, courtesy of D. Yong \citep[see][for details]{Yong05, Yong12}. We have McDonald 2.7m spectra of five stars in NGC 7789 courtesy of Michael Briley (private communication, 2013), and VLT UVES spectra of four stars in NGC 6192 from Laura Magrini \citep[see][for details]{Magrini10}. We have recently taken additional spectra of two stars each in NGC 1817, NGC 2141, and NGC 6939 with the Apache Point Observatory (APO) 3.5m (see below). We have 23 OCs total in our sample, which are summarized in Table \ref{tab:clusters} along with parameters, references, and sources of data. The signal-to-noise of these data range from 60 to 180 per pixel, and the resolution ranges from R$\sim$25,000 (older KPNO data) to R$\sim$47,000 (UVES, Keck, and McDonald data).
\\
\indent The \citet{Salaris04} ages were used for all clusters where available \citep[NGC 6192 was not included in their sample; we use an age of 0.18 Gyr from][]{Claria06} except for Berkeley 31 and Collinder 261; these two clusters have seen significant variations in calculated ages. The first photometric study of Be 31 produced an age of 8 Gyr, which was steadily revised downwards over time to $\sim$2.5 Gyr with a corresponding increase in R$_{\mathrm{GC}}$ from $\sim$13 to 16 kpc \citep{Guetter93, JP94, Salaris04, Hasegawa04, Cignoni11}. We adopt the most recent age and galactocentric distance determinations, 2.6 Gyr, and 16.3 kpc, from \citet{Cignoni11}. Cr 261's calculated age has similarly decreased over time from $\sim$10 to 6 Gyr, although its galactocentric radius has remained around 7.5 kpc \citep{JP94, Carraro98, Salaris04, BT06}. Again, we adopt the most recently determined parameters from \citet{BT06}.
\\
\indent Be 31 also presents a challenge in membership determination. \citet{Friel02} present low-resolution spectroscopic data for 17 stars in the field of Be 31, and find a mean cluster velocity of +61 $\pm$ 15 km s$^{-1}$. \citet{Yong05} find a similar cluster radial velocity of +60 $\pm$ 10 km s$^{-1}$ for five stars in the field of Be 31, though they note the large scatter and that the CMD is complex and may have significant contamination from field stars. They analyze star 886, with a radial velocity of +56.6 km s$^{-1}$, for abundances. \citet{Friel10} observed two other stars in the field of Be 31 (260 and 1295) and found them to have velocities that differed by 19 km s$^{-1}$, neither of which agreed with that of star 886 from \citet{Yong05}. In an effort to resolve the membership status of these stars, Friel (private communication, 2015) has observed a sample of 57 stars in the field of Be 31 with the Hydra multi-object spectrograph at WIYN\footnote{The WIYN Observatory is a joint facility of the University of Wisconsin-Madison, Indiana University, the National Optical Astronomy Observatory and the University of Missouri.}. These observations show a quite broad velocity distribution, but with a clear signature of the cluster velocity at $\sim$55 km s$^{-1}$. This cluster velocity indicates that the two stars observed by \citet{Friel10}, are in fact, not members of the cluster, while the velocity of star 886 is coincident with the cluster mean velocity. We therefore restrict our analysis of Be 31 to star 886.

\floattable
\begin{deluxetable}{l r r r r r r r r}
\tabletypesize{\scriptsize}
\tablewidth{0pt}
\tablecolumns{9}
\tablecaption{Clusters in the Sample \label{tab:clusters}}
\tablehead{\colhead{Cluster} & \colhead{l} & \colhead{b} & \colhead{d$_{\odot}$} & \colhead{Age} & \colhead{R$_{\mathrm{GC}}$\tablenotemark{a}} & \colhead{Ref.\tablenotemark{b}} & \colhead{No. of} & \colhead{Telescope} \\ \colhead{} & \colhead{(deg.)} & \colhead{(deg.)} & \colhead{(kpc)} & \colhead{(Gyr)} & \colhead{(kpc)} & \colhead{} & \colhead{Stars} & \colhead{}}
\startdata
Be 17 & 175.7 & -3.7 & 2.7 & 10.1 & 11.2 & F05 & 3 & KPNO 4m \\
Be 18 & 163.6 & 5.0 & 5.4 & 5.7 & 13.7 & Y12 & 2 & Keck \\
Be 21 & 186.8 & -2.5 & 6.2 & 2.2 & 14.7 & Y12 & 2 & Keck \\
Be 22 & 199.9 & -8.1 & 6.2 & 4.3 & 14.4 & Y12 & 2 & Keck \\
Be 31\tablenotemark{c} & 206.2 & 5.1 & 8.3 & 2.6 & 16.3 & C11 & 1 & CTIO 4m \\
Be 32 & 208.0 & 4.4 & 3.4 & 5.9 & 11.6 & FJP10 & 2, 2 & KPNO 4m, Keck \\
Be 39 & 223.5 & 10.1 & 4.3 & 7.0 & 11.9 & FJP10 & 4 & KPNO 4m \\
Cr 261\tablenotemark{c} & 301.7 & -5.6 & 2.7 & 6.0 & 7.4 & BT06 & 2 & CTIO 4m \\
M67 & 215.6 & 31.7 & 0.9 & 4.3 & 9.1 & FJP10 & 3 & KPNO 4m \\
NGC 188 & 122.8 & 22.5 & 1.7 & 6.3 & 9.4 & FJP10 & 4 & KPNO 4m \\
NGC 1193 & 146.8 & -12.2 & 5.8 & 4.2 & 13.6 & FJP10 & 1 & HET \\
NGC 1245 & 146.6 & -8.9 & 3.0 & 1.1 & 11.1 & J11 & 4 & KPNO 4m \\
NGC 1817 & 186.1 & -13.1 & 1.5 & 1.1 & 10.0 & J09 & 2, 2 & KPNO 4m, APO 3.5m \\
NGC 1883 & 163.1 & 6.2 & 3.9 & 0.7 & 12.3 & J09 & 2 & KPNO 4m \\
NGC 2141 & 198.0 & -5.8 & 3.9 & 2.4 & 12.2 & J09 & 2, 2 & KPNO 4m, APO 3.5m \\
NGC 2158 & 186.6 & 1.8 & 4.0 & 1.9 & 12.5 & J09 & 1 & KPNO 4m \\
NGC 2194 & 197.3 & -2.3 & 1.9 & 0.9 & 10.3 & J11 & 2 & KPNO 4m \\
NGC 2355 & 203.4 & 11.8 & 1.9 & 0.8 & 10.2 & J11 & 3 & KPNO 4m \\
NGC 6192\tablenotemark{c} & 340.7 & 2.1 & 1.5 & 0.2 & 7.1 & C06 & 4 & VLT \\
NGC 6939 & 95.9 & 12.3 & 1.8 & 1.2 & 8.9 & A04 & 4, 2 & KPNO 4m, APO 3.5m \\
NGC 7142 & 105.0 & 9.0 & 1.9 & 4.0 & 9.2 & J08 & 4 & KPNO 4m \\
NGC 7789 & 115.5 & -5.4 & 2.2 & 1.8 & 9.6 & J11 & 5 & McDonald 2.7m \\
PWM 4 & 116.0 & 0.3 & 7.2 & 7.0 & 13.3 & Y12 & 1 & Keck \\
\enddata
\tablenotetext{a}{R$_{\mathrm{GC},\odot}$ = 8.5 kpc}
\tablenotetext{b}{References for distances: C11 = \citet{Cignoni11}; BT06 = \citet{BT06}; C06 = \citet{Claria06}; F05 = \citet{Friel05}; FJP10 = \citet{Friel10}; J08 = \citet{Jacobson08}; J09 = \citet{Jacobson09}; J11 = \citet{Jacobson11}; Y05 = \citet{Yong05}; Y12 = \citet{Yong12}}
\tablenotetext{c}{Cluster age not taken from \citet{Salaris04} (see text for details)}
\label{clusters}
\end{deluxetable}

\subsection{New APO Data}  \label{sec:APO}

\indent We obtained additional APO spectra for several previously observed clusters to better define the dispersions in these clusters, which were found to be unusually large for some elements in \citet{JF13}. Our target selection for the new APO observations of NGC 1817 and NGC 6939 was based on WIYN 3.5m observations of the two clusters presented in \citet{Jacobson11} and \citet{Jacobson07}, respectively, which confirmed these stars as members based on radial velocity. Our two targets in NGC 2141 were selected based on radial velocity confirmation from \citet{Yong05}. Our APO data were taken on the nights of 16 Nov. 2014 (NGC 1817 and NGC 6939), 6, 15, 26, and 28 of Jan. 2015 (NGC 2141). We used the ARC Echelle Spectrograph on standard settings (2048x2048 SITe CCD, 3200-10000$\mathrm{\AA}$, R $\sim$ 32000) to obtain spectra of two stars per cluster. Because we targeted red giant stars, our signal-to-noise was unfortunately limited in the blue, but we were able to measure lines to $\sim$4500$\mathrm{\AA}$ for each spectrum. Data were reduced using the ARCES reduction guide by Julie Thorburn\footnote{http://www.apo.nmsu.edu/arc35m/Instruments/\\ARCES/images/echelle\_data\_reduction\_guide.pdf}. The signal-to-noise of our combined and reduced spectra at 6000$\mathrm{\AA}$ ranges from 75 to 90. Details for these observations are summarized in Table \ref{tab:APO}.
\\
\indent Radial velocities for our APO spectra were measured with FXCOR in IRAF\footnote{IRAF is distributed by the National Optical Astronomy Observatory, which is operated by the Association of Universities for Research in Astronomy, Inc., under cooperative agreement with the National Science Foundation.} by measuring a radial velocity for each aperture, then taking the weighted mean of the aperture velocities and errors. However, the error on the weighted mean of these values significantly underestimates the errors caused by night-to-night zero point variations. For the star we observed over three nights, N2141 514, we find that measured heliocentric radial velocities for different nights have a dispersion of 0.6 km~s$^{-1}$, so we adopt this as the error on our velocities. We generally find good agreement within the errors for available literature radial velocities for N1817 and N2141 stars; these have differences less than 2 km~s$^{-1}$ with \citet{Mermilliod03} and \citet{Jacobson11} for N1817 and \citet{Yong05} for N2141. However, our N6939 velocities are lower than those found in \citet{Milone94} and \citet{Jacobson07} by several km~s$^{-1}$. It is possible that both stars may be binaries, or this may be due to differences in zero-point corrections, but our measured radial velocities are close enough to those in the literature that we are confident these stars are cluster members. It is also worth noting that \citet{Mermilliod03} classify N1817 1456 as a possible spectroscopic binary due to the significant dispersion in CORAVEL radial velocity measurements, but we do not see any binarity in the line profiles of our spectrum.

\floattable
\begin{deluxetable}{l r r r r r r l r r r r}
\tabletypesize{\scriptsize}
\tablewidth{0pt}
\tablecolumns{12}
\tablecaption{APO Spectroscopic Observations \label{tab:APO}}
\tablehead{\colhead{} & \colhead{} & \colhead{$\alpha_{J2000}$} & \colhead{$\delta_{J2000}$} & \colhead{} & \colhead{Phot.} & \colhead{Exp. UT} & \colhead{Exp. } & \colhead{Lit. V$_r$} & \colhead{V$_r$} & \colhead{V$_r$} & \colhead{} \\ \colhead{Cluster} & \colhead{Star} & \colhead{(h:m:s)} & \colhead{(d:m:s)} & \colhead{V} & \colhead{Ref.\tablenotemark{a}} & \colhead{Date} & \colhead{Time} & \colhead{(km~s$^{-1}$)} & \colhead{Ref.\tablenotemark{a}} & \colhead{(km~s$^{-1}$)} & \colhead{S/N}}
\startdata
N1817 & 206 & 05:13:01.80 & 16:41:14.4 & 11.92 & M03 & 2014 11 16 & 5 x 600s & 65.5 $\pm$ 0.4 & M03 & 64.4 & 79 \\
N1817 & 1456 & 05:12:32.83 & 16:28:25.3 & 11.43 & M03 & 2014 11 16 & 5 x 600s & 66.0 $\pm$ 1.0 & M03 & 64.7 &  79 \\
N2141 & 514 & 06:03:00.00 & 10:32:24.0 & 14.09 & Y05 & 2015 01 07 & 3 x 3000s & 23.3 $\pm$ 0.9 & Y05 & 23.2 & 76 \\
 &  &  &  &  &  & 2015 01 16 & 3 x 3000s &  &  & 24.1 &  \\
 &  &  &  &  &  & 2015 01 27 & 1 x 3000s &  &  & 24.3 &  \\
N2141 & 1821 & 06:03:07.60 & 10:26:48.0 & 14.13 & Y05 & 2015 01 27 & 2 x 3000s & 24.8 $\pm$1.0 & Y05 & 25.3 & 76 \\ 
 &  &  &  &  &  & 2015 01 29 & 5 x 3000s &  &  & 25.4 &  \\
N6939 & 53 & 20:31:01.90 & 60:38:11.6 & 12.78 & MN07 & 2014 11 16 & 4 x 1800s & -20.1 $\pm$ 0.7 & M94 & -23.6 & 93 \\
N6939 & 65 & 20:31:05.98 & 60:42:13.8 & 12.68 & MN07 & 2014 11 16 & 4 x 1800s & -18.6 $\pm$ 0.6 & M94 & -20.7 & 90 \\
\enddata
\tablenotetext{a}{References for literature photometry and/or radial velocities: M03 = \citet{Mermilliod03}; MN07 = \citet{MN07}; M94 = \citet{Milone94}; Y05 = \citet{Yong05}}
\end{deluxetable}

\section{Atmospheric Parameters and Iron}  \label{sec:params}

\indent Atmospheric parameters and Fe abundances for all stars described in \citet{JF13} are taken from that work so we will not describe them here. In order to remain consistent with our internal abundance scale, we have calculated our own atmospheric parameters and iron abundances for all stars we have not previously analyzed. We used spherical MARCS\footnote{Retrieved from http://marcs.astro.uu.se/} model atmospheres \citep{Gustafsson08} with the 2010 version of MOOG \citep{Sneden73} for all abundance calculations. We selected a set of relatively unblended iron lines from the Gaia-ESO Survey (GES) to measure equivalent widths \citep{Heiter15}; Fe lines, atomic data, and equivalent widths are listed in Table \ref{tab:linelist}. 
\\
\indent Echelle [Fe/H] abundances taken from \citet{JF13} were calculated using a linelist with Arcturus-based log(gf) values, different than the GES linelist we present here for the newest stellar abundance measurements. In a previous abundance study of NGC 7789 using lower-resolution WIYN Hydra spectra, we demonstrated that the differences between the two linelists are small and do not give rise to any systematic differences in atmospheric parameters \citep{Overbeek15}. We did find that for the NGC 7789 data, using the Arcturus-based linelist resulted in a $\sim$0.1 dex higher [Fe/H], but comparing the three clusters for which we have iron abundances based on both linelists (NGC 1817, NGC 2141, and NGC 6939) we find no difference in the cluster [Fe/H] determined with the Arcturus-based vs. the GES log(gf) values.
\\
\indent Starting from the literature atmospheric parameters for Be 31 and NGC 6192, or typical previously determined clump parameters for NGC 7789, NGC 1817, NGC 2141, and NGC 6939, we determined parameters spectroscopically with the typical method: adjusting temperature to minimize abundance trends with excitation potential, adjusting microturbulent velocity to minimize abundance trends with reduced equivalent width, adjusting the gravity to match abundances from Fe I and Fe II, and repeating until a good solution is found. We have also re-calculated spectroscopic parameters for the two warmer Cr 261 stars from \citet{Friel03} since the Fe linelist used here is significantly different; setting the continuum was difficult for the cooler Cr 261 stars and resulted in large abundance errors, so we do not include them in this sample. Derived atmospheric parameters and Fe abundances can be found in Table \ref{tab:atm_params}, along with literature parameters (where available) for comparison.

\floattable
\begin{deluxetable}{l r r r r r r r r r r h h h h h h h h h h h}
\tabletypesize{\scriptsize}
\tablewidth{0pt}
\tablecolumns{7}
\tablecaption{Fe Linelist Parameters and Equivalent Width Measurements\tablenotemark{a} \label{tab:linelist}}
\tablehead{\colhead{$\lambda$} & \colhead{Sp.} & \colhead{E.P.} & \colhead{log(gf)} & \colhead{Be31 886} & \colhead{Cr261 1045} & \colhead{Cr261 1080} & \colhead{N1817 206} & \colhead{N1817 1456} & \colhead{N2141 514} & \colhead{N2141 1821} & \nocolhead{N6192 9} & \nocolhead{N6192 45} & \nocolhead{N6192 96} & \nocolhead{N6192 137} & \nocolhead{N6939 53} & \nocolhead{N6939 65} & \nocolhead{N7789 212} & \nocolhead{N7789 468} & \nocolhead{N7789 605} & \nocolhead{N7789 765} & \nocolhead{N7789 958} \\ colhead{$(\mathrm{\AA})$} & \colhead{} & \colhead{(eV)} & \colhead{} & \colhead{EQW (m$\mathrm{\AA})$} & \colhead{EQW (m$\mathrm{\AA})$} & \colhead{EQW (m$\mathrm{\AA})$} & \colhead{EQW (m$\mathrm{\AA})$} & \colhead{EQW (m$\mathrm{\AA})$} & \colhead{EQW (m$\mathrm{\AA})$} & \colhead{EQW (m$\mathrm{\AA})$} & \nocolhead{EQW (m$\mathrm{\AA})$} & \nocolhead{EQW (m$\mathrm{\AA})$} & \nocolhead{EQW (m$\mathrm{\AA})$} & \nocolhead{EQW (m$\mathrm{\AA})$} & \nocolhead{EQW (m$\mathrm{\AA})$} & \nocolhead{EQW (m$\mathrm{\AA})$} & \nocolhead{EQW (m$\mathrm{\AA})$} & \nocolhead{EQW (m$\mathrm{\AA})$} & \nocolhead{EQW (m$\mathrm{\AA})$} & \nocolhead{EQW (m$\mathrm{\AA})$} & \nocolhead{EQW (m$\mathrm{\AA})$} }
\startdata
5242.49 & 26.0 & 3.63 & -0.967 & $\ldots$ & $\ldots$ & $\ldots$ & 122.6 & 128.1 & 130.9 & 117.8 & 140.1 & 128.6 & 143.7 & 150.2 & 120.8 & 119.9 & 131.9 & 149.3 & 133.0 & 155.0 & 123.3 \\
5365.40 & 26.0 & 3.57 & -1.020 & $\ldots$ & $\ldots$ & 111.3 & 127.4 & $\ldots$ & $\ldots$ & $\ldots$ & 140.4 & 129.0 & $\ldots$ & 144.8 & $\ldots$ & 121.6 & $\ldots$ & $\ldots$ & $\ldots$ & $\ldots$ & $\ldots$ \\
5379.57 & 26.0 & 3.69 & -1.514 & $\ldots$ & 92.5 & 93.9 & 106.5 & 104.8 & 110.7 & 98.3 & 113.3 & 104.3 & 121.4 & 125.9 & 97.5 & 97.5 & 116.0 & 115.5 & 99.1 & 107.6 & 96.7 \\
5417.03 & 26.0 & 4.42 & -1.580 & $\ldots$ & 67.8 & 57.2 & 57.3 & 68.6 & 62.2 & 51.3 & 77.4 & 66.8 & 75.3 & 80.8 & 66.1 & 64.2 & 71.9 & 73.3 & 71.9 & 76.0 & 66.8 \\
5466.99 & 26.0 & 3.57 & -2.233 & $\ldots$ & 79.3 & 73.3 & 77.0 & 88.8 & 85.1 & 76.9 & 89.1 & 82.1 & 92.4 & 103.3 & 81.0 & 72.0 & 97.8 & $\ldots$ & 86.3 & 103.5 & 78.0 \\
5633.95 & 26.0 & 4.99 & -0.230 & 76.0 & 89.6 & 84.9 & 90.1 & 88.4 & 92.9 & 84.0 & 107.2 & 91.5 & 111.3 & 113.5 & 91.1 & 83.5 & 96.4 & 103.7 & 102.6 & 104.1 & 80.7 \\
5662.52 & 26.0 & 4.18 & -0.447 & 102.6 & 118.2 & 114.8 & 123.5 & 129.5 & $\ldots$ & 115.9 & 144.1 & 127.4 & $\ldots$ & 150.9 & 116.7 & 116.0 & 141.2 & 139.7 & 132.4 & 134.8 & 126.2 \\
5701.54 & 26.0 & 2.56 & -2.193 & 117.5 & 143.7 & 138.3 & 132.7 & 139.0 & 158.3 & 135.0 & 153.7 & 146.5 & 164.8 & $\ldots$ & 131.8 & 132.4 & $\ldots$ & $\ldots$ & $\ldots$ & $\ldots$ & $\ldots$ \\
5705.46 & 26.0 & 4.30 & -1.355 & 70.3 & 73.8 & 77.2 & 70.9 & 79.6 & 82.9 & 75.4 & 86.4 & 77.2 & 96.2 & 89.0 & 76.4 & 75.6 & $\ldots$ & $\ldots$ & $\ldots$ & $\ldots$ & $\ldots$ \\
5753.12 & 26.0 & 4.26 & -0.623 & 86.0 & 104.9 & 102.5 & 102.8 & 117.1 & 111.9 & 108.0 & 131.2 & 117.7 & 133.8 & 133.6 & 119.0 & 113.8 & $\ldots$ & $\ldots$ & $\ldots$ & $\ldots$ & $\ldots$ \\
\enddata
\tablenotetext{a}{Table \ref{tab:linelist} is published in its entirety in the electronic edition of the AJ. A portion is shown here for guidance regarding its form and content.}
\end{deluxetable}

\floattable
\begin{deluxetable}{l r c r r c r r r r r r}
\tabletypesize{\scriptsize}
\tablewidth{0pt}
\tablecolumns{12}
\tablecaption{Atmospheric Parameters of Newly Analyzed Stars \label{tab:atm_params}}
\tablehead{\colhead{Cluster} & \colhead{Star} & \colhead{Source\tablenotemark{b}} & \colhead{T$_{\mathrm{eff}}$} & \colhead{log(g)} & \colhead{$\xi$} & \colhead{FeI} & \colhead{$\sigma$FeI} & \colhead{FeI} & \colhead{FeII} & \colhead{$\sigma$FeII} & \colhead{FeII} \\ \colhead{} & \colhead{I.D.} & \colhead{} & \colhead{(K)} & \colhead{(dex)} & \colhead{(km s$^{-1}$)} & \colhead{} & \colhead{(dex)} & \colhead{no.} & \colhead{} & \colhead{(dex)} & \colhead{no.}}
\startdata
Be 31 & 886 & This work & 4500 & 1.8 & 1.05 & 7.17 & 0.12 & 55 & 7.18 & 0.04 & 7 \\
Cr 261 & 1045 & This work & 4500 & 2.0 & 1.15 & 7.49 & 0.13 & 60 & 7.51 & 0.13 & 10 \\
Cr 261 & 1080 & This work & 4500 & 2.1 & 1.10 & 7.44 & 0.13 & 61 & 7.42 & 0.08 & 10 \\
N1817 & 206 & This work & 4750 & 2.1 & 1.35 & 7.45 & 0.12 & 60 & 7.45 & 0.13 & 11 \\
N1817 & 1456 & This work & 4750 & 2.1 & 1.45 & 7.48 & 0.12 & 58 & 7.48 & 0.10 & 11 \\
N2141 & 514 & This work & 4400 & 1.6 & 1.35 & 7.44 & 0.12 & 59 & 7.42 & 0.09 & 11 \\
N2141 & 1821 & This work & 4350 & 1.5 & 1.05 & 7.44 & 0.13 & 59 & 7.44 & 0.12 & 10 \\
N6192 & 9 & This work & 4900 & 1.9 & 1.60 & 7.69 & 0.09 & 53 & 7.69 & 0.11 & 11 \\
N6192 & 45 & This work & 4900 & 2.5 & 1.35 & 7.61 & 0.11 & 55 & 7.62 & 0.06 & 11 \\
N6192 & 96 & This work & 4800 & 1.9 & 1.75 & 7.57 & 0.12 & 52 & 7.58 & 0.07 & 11 \\
N6192 & 137 & This work & 4550 & 1.8 & 1.65 & 7.59 & 0.13 & 47 & 7.61 & 0.09 & 11 \\
N6939 & 53 & This work & 4900 & 2.4 & 1.30 & 7.56 & 0.11 & 59 & 7.56 & 0.12 & 11 \\
N6939 & 65 & This work & 4950 & 2.3 & 1.40 & 7.50 & 0.11 & 57 & 7.49 & 0.10 & 11 \\
N7789 & 212 & This work & 4400 & 1.8 & 1.45 & 7.50 & 0.15 & 56 & 7.50 & 0.17 & 14 \\
N7789 & 468 & This work & 4250 & 1.5 & 1.65 & 7.46 & 0.14 & 54 & 7.48 & 0.15 & 14 \\
N7789 & 605 & This work & 4900 & 2.5 & 1.55 & 7.52 & 0.16 & 58 & 7.53 & 0.11 & 12 \\
N7789 & 765 & This work & 4450 & 2.0 & 1.50 & 7.54 & 0.14 & 57 & 7.55 & 0.13 & 13 \\
N7789 & 958 & This work & 5050 & 2.8 & 1.50 & 7.56 & 0.16 & 59 & 7.55 & 0.10 & 13 \\
\tableline
Be 31 & 886 & Y05 & 4490 & 1.9 & 1.22 & 6.97 & 0.23 & 35 & 7.05 & 0.18 & 7 \\
Cr 261 & 1045 & C05 & 4470 & 2.07 & 1.23 & 7.55 & 0.16 & 126 & 7.49 & 0.19 & 10 \\
Cr 261 & 1045 & F03 & 4400 & 1.5 & 1.20 & 7.36 & 0.20 & 37 & 7.31 & 0.18 & 6 \\
Cr 261 & 1080 & C05 & 4500 & 2.09 & 1.23 & 7.54 & 0.15 & 114 & 7.48 & 0.26 & 11 \\
Cr 261 & 1080 & F03 & 4490 & 2.2 & 1.20 & 7.41 & 0.25 & 41 & 7.45 & 0.22 & 7\\
N6192 & 9 & M10 & 5050 & 2.3 & 1.75 & 7.66 & 0.07 & 59 & $\ldots$ & $\ldots$ & 8 \\
N6192 & 45 & M10 & 5020 & 2.55 & 1.60 & 7.55 & 0.08 & 119 & $\ldots$ & $\ldots$ & 9 \\
N6192 & 96 & M10 & 5050 & 2.3 & 2.10 & 7.60 & 0.10 & 77 & $\ldots$ & $\ldots$ & 8 \\
N6192 & 137 & M10 & 4670 & 2.1 & 1.80 & 7.54 & 0.08 & 91 & $\ldots$ & $\ldots$ & 9 \\
N6939 & 53 & JFP07 & 4920 & 2.4 & 1.70 & 7.41 & 0.17 & 22 & 7.54 & 0.10 & 7 \\
N6939 & 65 & JFP07 & 5100 & 3.0 & 1.80 & 7.55 & 0.20 & 22 & 7.56 & 0.06 & 8 \\
N7789 & 468 & T05 & 4190 & 1.3 & 1.90 & 7.44\tablenotemark{a} & 0.10 & 22 & 7.44\tablenotemark{a} & 0.03 & 3 \\
N7789 & 765 & T05 & 4430 & 1.8 & 1.70 & 7.48\tablenotemark{a} & 0.09 & 22 & 7.48\tablenotemark{a} & 0.07 & 3 \\
\enddata
\tablenotetext{a}{Solar abundances are not given in \citet{Taut05}, so we assume $\epsilon$(Fe$_{\odot}$) = 7.52}
\tablenotetext{b}{References for atmospheric parameters: C05 = \citet{Carretta05}; F03 = \citet{Friel03}; JFP07 = \citet{Jacobson07}; M10 = \citet{Magrini10}; T05 = \citet{Taut05}; Y05 = \citet{Yong05}}
\end{deluxetable}

\indent Our atmospheric parameters for Be 31 match those in \citet{Yong05} well; the Fe abundance differs by 0.15 dex, within the errors for the two measurements. Parameters for Collinder 261 also match those in the literature, and Fe differences are again within the expected error. Our effective temperatures for NGC 6192 are systematically 100 K below \citet{Magrini10}, and our log(g) values are $\sim$0.3 dex lower, although Fe abundances align well. Parameters for two overlapping stars between our sample and \citet{Taut05} also have very similar parameters and abundances. Our microturbulent velocities are systematically low compared to all literature values by about 0.2 km s$^{-1}$.
\\
\indent Our derived Fe abundance for Be 31 compares well with literature values; the cluster has been consistently estimated to have a subsolar [Fe/H] $\sim$ -0.4 dex as estimated by photometry \citep{Guetter93, Hasegawa04} and low-resolution spectroscopy \citep{Friel02}. For Cr 261, \citet{Friel03} find slightly subsolar Fe abundances, but \citet{Carretta05}, \citet{deSilva07}, and \citet{Mishenina15} all find solar Fe abundances. Many previous spectroscopy and photometry studies have placed NGC 7789 at solar or slightly sub-solar metallicities \citep[e.g.,][]{Pilachowski85, Twarog97, Schonberner01, Friel02, Taut05, Pancino10, Overbeek15}. Though our sample size per cluster is smaller than many of these studies, we can place some confidence in our derived Fe abundances.
\\
\indent We determine atmospheric parameter errors by taking into account both internal uncertainties (due to residual slopes with excitation potential and reduced equivalent widths for temperatures and microturbulent velocities and differences between FeI and FeII abundances for gravities) and systematic differences between our parameters and literature values. The residual slopes in abundance with excitation potential are small, on the order of 0.005 dex eV$^{-1}$, and the typical uncertainty on the slopes is $\sim$0.015 dex eV$^{-1}$, which correspond to changes in temperatures of $\pm$20 K and $\pm$75 K for a star with an effective temperature of 4500 K. However, the typical difference between temperatures from this work and literature values in Table \ref{tab:atm_params} is about 100 K, so we estimate our temperature errors conservatively as $\pm$100 K. For microturbulent velocity, the typical slope and error on the slope of abundance with reduced equivalent width are 0.02 and 0.09, which correspond to differences in v.t. of 0.02 and 0.08 km s$^{-1}$, but the median difference between our values and literature values is 0.2 km s$^{-1}$, so we adopt 0.2 km s$^{-1}$ for the error on v.t. Differences between our Fe I and Fe II values are no more than $\pm$0.02 dex, corresponding to an internal uncertainty in log(g) of 0.1 dex, but again the differences between studies are larger, so we adopt the median difference between our and literature log(g) of 0.2 dex as the error on the gravities. Our [Fe I/H] dispersions are typically 0.10 - 0.15 dex, and Fe II dispersions are similar. We discuss the impact of these errors on neutron-capture abundances in Section \ref{sec:errors}.

\section{Neutron-Capture Abundance Measurements}  \label{sec:abuns}

\begin{figure*}
\epsscale{0.8}
\plotone{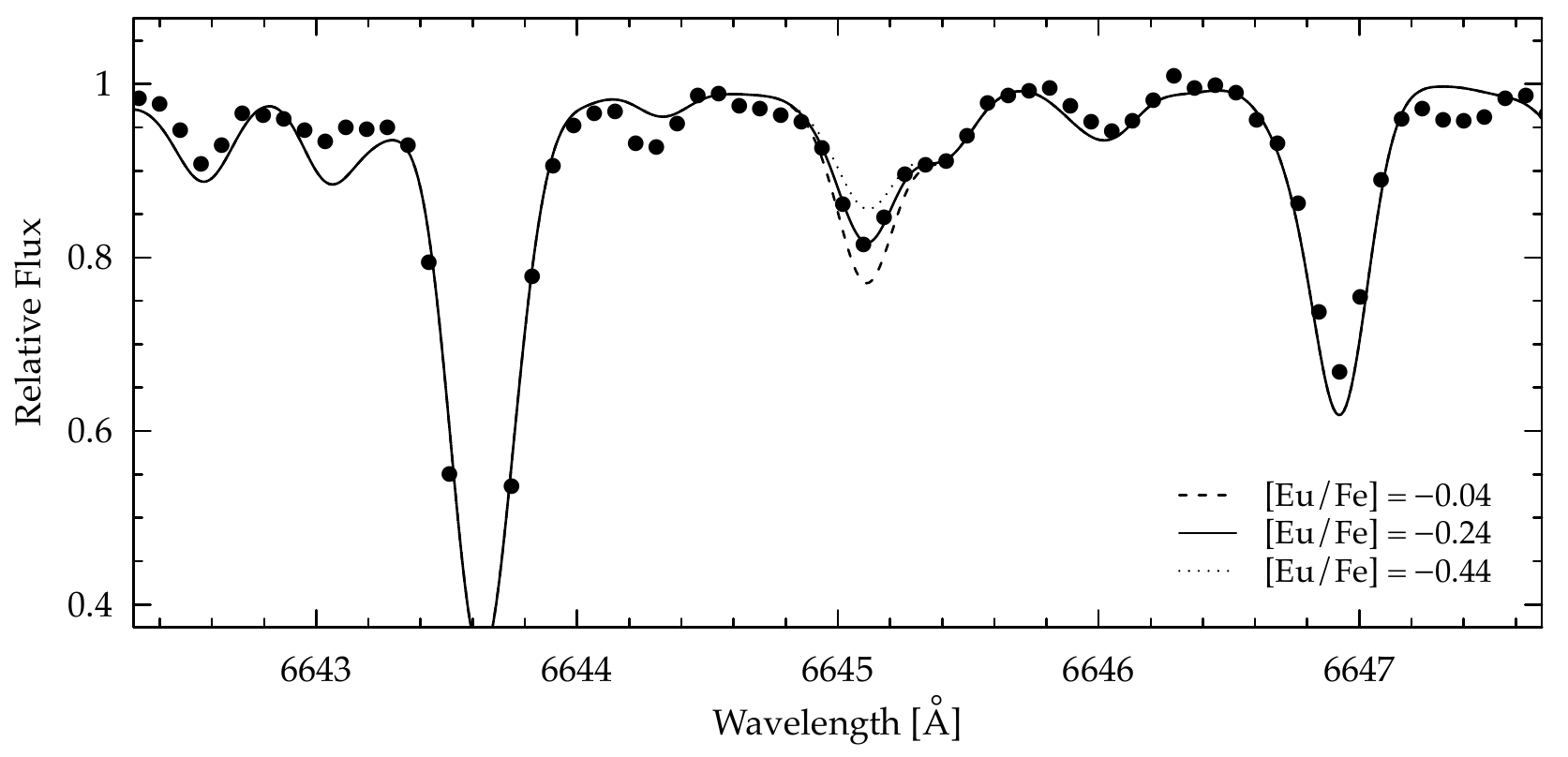}
\caption{Example fits to the 6645$\mathrm{\AA}$ europium line in NGC 6939 star 31. Dots are the observed spectrum, the solid line represents the selected best fit, and the dashed and dotted lines are the best fit abundance +/- 0.2 dex, respectively.}
\label{Eu_fit}
\end{figure*}

\begin{figure}
\epsscale{1.0}
\plotone{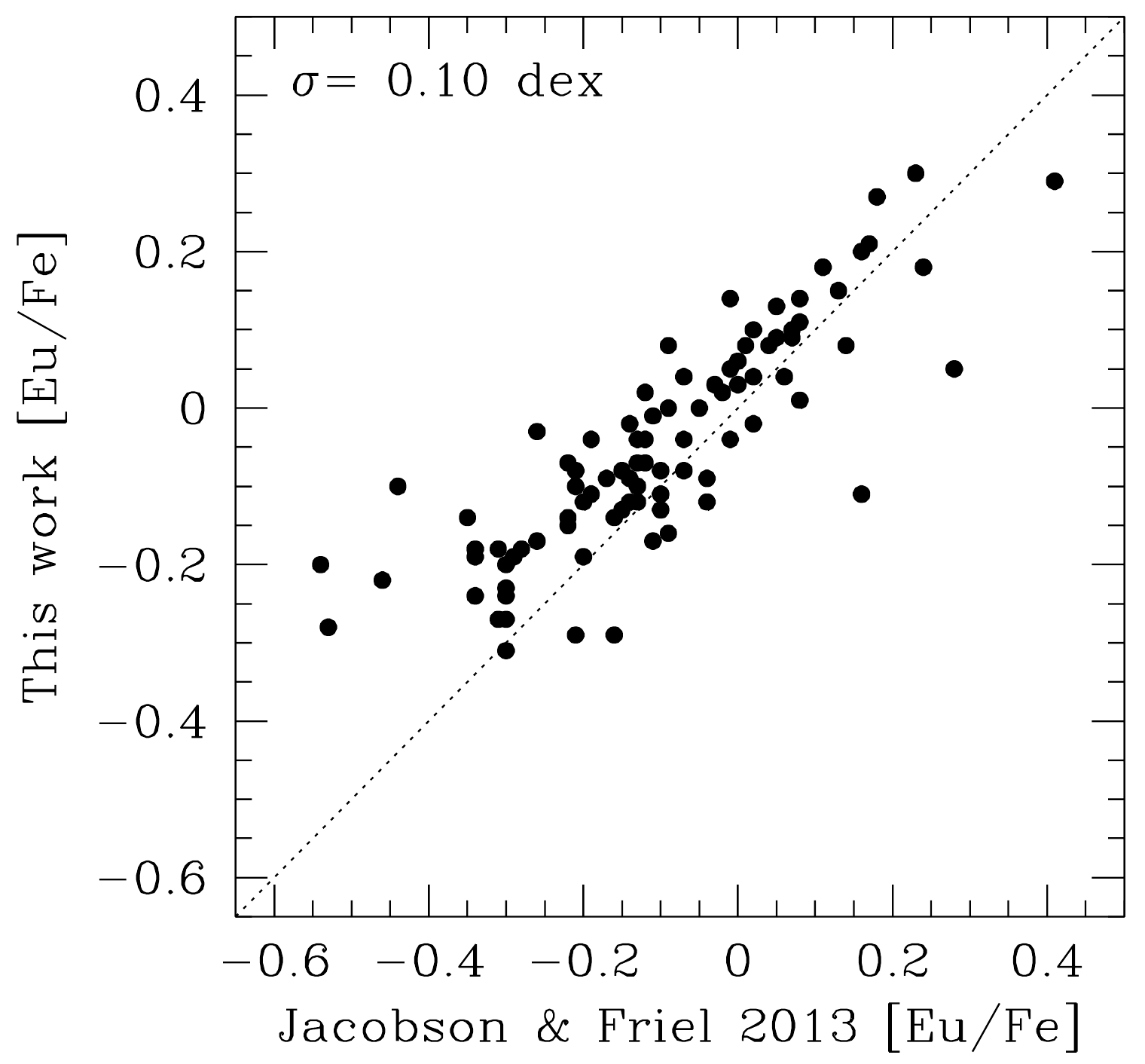}
\caption{Comparison between [Eu/Fe] values for the 6437 and 6645$\mathrm{\AA}$ lines as measured by \citet{JF13} and this work. The dashed line represents a 1:1 correspondence.}
\label{Eu_comp}
\end{figure}

\indent We used the MOOG LTE spectral synthesis program to measure all neutron-capture elements presented here \citep{Sneden73}. We have used previously determined Mg, Si, and Ti abundances (from our determinations where available; otherwise from literature sources described in Section \ref{sec:data}) for lines with significant blending with these elements. We have also assumed [C/Fe], [N/Fe], and [O/Fe] = -0.20, +0.20, and 0.00 dex, respectively, and a C$^{12}$/C$^{13}$ ratio of 10 for our sample stars, based on typical red clump star abundances from \citet{Taut10}. Smoothing for each instrument and wavelength region was set by carefully fitting unblended surrounding lines within a 10$\mathrm{\AA}$ window of the feature of interest.
\\
\indent In the interest of making all the abundances as homogenous as possible, we have employed a C code to automate the spectral synthesis. It takes a smoothing value, continuum parameter, and set of auxiliary abundances for blends, and calculates synthetic spectra for a range of abundances of the desired element using MOOG. It then compares the observed spectrum in the area of the feature being measured, and selects the best fit to the data based on the $\chi^2$ value. In Figure \ref{Eu_fit} we show a sample europium synthesis fit selected by the program for NGC 6939 star 31; the selected best fit abundance is marked with a solid line, and the best fit abundance plus and minus 0.2 dex are marked with dashed and dotted lines, respectively. As we show in Section \ref{sec:errors}, the error on the Eu abundance for this star due to atmospheric parameter and continuum errors (added in quadrature) is 0.2 dex.
\\
\indent Our automated synthesis program generally does well selecting the best fit. In Figure \ref{Eu_comp} we show the results of europium measurements from the automated synthesis versus previous europium results from synthesis of the 6437 and 6645$\mathrm{\AA}$ lines `by eye' \citep{JF13}. There is a slight offset between the two measurements: $<$[Eu/Fe]$_{\mathrm{JF13}}$ - [Eu/Fe]$_{\mathrm{this}}>$ = 0.05 $\pm$ 0.10 dex. In Figure \ref{Eu_comp} the largest positive differences ([Eu/Fe]$_{\mathrm{JF13}}$ $<$ [Eu/Fe]$_{\mathrm{this}}$) are for two relatively high-metallicity clusters (NGC 7142 and NGC 188), which could be explained by differences in the modeling of CN features or continuum setting. Over the range of the data, the difference from a slope of 1 (dotted line) is smaller than the scatter around the fit, so we conclude that there is no systematic difference between fits by eye and by our automated synthesis program.
\\
\indent As we discuss in detail in later sections, stellar abundance dispersions are typically less than 0.10 dex, so the errors in the fits to individual features due to the S/N of the spectra should not be significantly larger than this. We discuss the possible effects of blending to various lines in the sections for each element, and the errors due to the continuum setting parameter are discussed in Section \ref{sec:errors}.
\\
\indent Linelists used for the synthesis of neutron-capture features were selected from recent literature studies of atomic data for each element. The selection of sources for individual lines is described in more detail in sections 4.1 - 4.6. Atomic data for blends and surrounding lines within the 10$\mathrm{\AA}$ fitting window were taken from a comprehensive linelist we have received courtesy of Chris Sneden. All linelists used for synthesis include molecular features. In order to establish the zero point for our abundances, some of which have little data in the literature, we have measured solar and Arcturus abundances for the selected neutron-capture lines using high-resolution and signal-to-noise spectra of \citet{Hinkle00}. For our solar synthesis, we use \citet{AG89} input abundances; for Arcturus, we use CNO abundances from \citet{Abia12} because these give the best visual fit to molecular features in our linelists, and other elemental abundances are taken from \citet{RAP11}. The following sections discuss the solar abundances, hyperfine structure, isotopic ratios, and other atomic data for the individual elements we discuss in this paper. The wavelength and species of lines measured, type and sources of atomic data, and solar and Arcturus abundance measurements for each line are given in Table \ref{tab:solar_abs}. The Arcturus abundance ratios to Fe given in the last column are relative to the literature solar values from \citet{AG89}, not our measured solar abundances.

\floattable
\begin{deluxetable}{l c c c r c c c c r}
\tabletypesize{\scriptsize}
\tablewidth{0pt}
\tablecolumns{10}
\tablecaption{Solar and Arcturus Abundances \label{tab:solar_abs}}
\tablehead{\colhead{$\lambda$} & \colhead{Species} & \colhead{Structure?} & \colhead{E.P.} & \colhead{log(gf)\tablenotemark{a}} & \colhead{Source\tablenotemark{b}} & \colhead{A\&G89} & \colhead{Our} & \colhead{Our} & \colhead{Arc.} \\ \colhead{$[\mathrm{\AA}]$} & \colhead{} & \colhead{} & \colhead{} & \colhead{(dex)} & \colhead{} & \colhead{$\epsilon_{\odot}$} & \colhead{$\epsilon_{\odot}$} & \colhead{$\epsilon_{\mathrm{Arc.}}$} & \colhead{[X/Fe]\tablenotemark{c}}}
\startdata
5570.44 & Mo I & $\ldots$ & 1.335 & -0.34 & WB88 & 1.92 & 1.92\tablenotemark{d} & 1.58 & 0.19 \\
5751.41 & Mo I & $\ldots$ & 1.420 & -1.01 & WB88 & 1.92 & 1.92\tablenotemark{d} & 1.53 & 0.14 \\
6030.64 & Mo I & $\ldots$ & 1.531 & -0.52 & WB88 & 1.92 & 1.91 & 1.55 & 0.16 \\
5219.04 & Pr II & hfs & 0.795 & -0.05 & S09 & 0.71 & 0.66 & 0.18 & 0.00 \\
5259.72 & Pr II & hfs & 0.633 & 0.11 & S09 & 0.71 & 0.71 & 0.22 & 0.04 \\
5322.77 & Pr II & hfs & 0.482 & -0.12 & S09 & 0.71 & 0.63 & 0.14 & -0.04 \\
5352.40 & Pr II & hfs & 0.482 & -0.74 & S09 & 0.71 & 0.85 & 0.20 & 0.02 \\
5132.33 & Nd II & $\ldots$ & 0.559 & -0.71 & DH03 & 1.50 & 1.50\tablenotemark{d} & 1.03 & 0.06 \\
5306.46 & Nd II & $\ldots$ & 0.859 & -0.97 & DH03 & 1.50 & 1.57 & 1.04 & 0.07 \\
5356.96 & Nd II & $\ldots$ & 1.263 & -0.28 & DH03 & 1.50 & 1.46 & 0.94 & -0.03 \\
5416.37 & Nd II & $\ldots$ & 0.859 & -0.93 & M75 & 1.50 & 1.42 & 0.89 & -0.08 \\
5431.52 & Nd II & $\ldots$ & 1.120 & -0.47 & DH03 & 1.50 & 1.59 & 1.01 & 0.04 \\
5485.70 & Nd II & $\ldots$ & 1.263 & -0.12 & DH03 & 1.50 & 1.61 & 1.01 & 0.04 \\
5533.82 & Nd II & $\ldots$ & 0.559 & -1.23 & DH03 & 1.50 & 1.41 & 0.77 & -0.20 \\
5740.86 & Nd II & $\ldots$ & 1.159 & -0.53 & DH03 & 1.50 & 1.46 & 0.95 & -0.02 \\
5811.57 & Nd II & $\ldots$ & 0.859 & -0.86 & DH03 & 1.50 & 1.43 & 0.90 & -0.07 \\
6385.15 & Nd II & $\ldots$ & 1.160 & -0.77 & M75 & 1.50 & 1.41 & 0.87 & -0.10 \\
4129.72 & Eu II & hfs/iso & 0.000 & 0.22 & L01 & 0.51 & 0.50 & 0.27 & 0.29 \\
4205.04 & Eu II & hfs/iso & 0.000 & 0.21 & L01 & 0.51 & 0.56 & 0.32 & 0.34 \\
6437.64 & Eu II & hfs/iso & 1.319 & -0.32 & L01 & 0.51 & 0.51 & 0.26 & 0.28 \\
6645.10 & Eu II & hfs/iso & 1.379 & 0.12 & L01 & 0.51 & 0.49 & 0.25 & 0.27 \\
4085.56 & Gd II & $\ldots$ & 0.731 & -0.01 & DH06 & 1.12 & 1.15 & 0.97 & 0.38 \\
4191.08 & Gd II & $\ldots$ & 0.427 & -0.48 & DH06 & 1.12 & 1.14 & 1.07 & 0.48 \\
4316.05 & Gd II & $\ldots$ & 0.662 & -0.45 & DH06 & 1.12 & 1.12\tablenotemark{d} & 0.97 & 0.38 \\
4483.33 & Gd II & $\ldots$ & 1.059 & -0.42 & DH06 & 1.12 & 1.17 & 0.96 & 0.37 \\
4498.29 & Gd II & $\ldots$ & 0.427 & -1.08 & DH06 & 1.12 & 1.12\tablenotemark{d} & 0.96 & 0.37 \\
4073.12 & Dy II & $\ldots$ & 0.538 & -0.32 & S09 & 1.10 & 1.17 & 0.97 & 0.40 \\
4449.70 & Dy II & $\ldots$ & 0.000 & -1.03 & S09 & 1.10 & 1.10 & 0.98 & 0.41 \\
5169.69 & Dy II & $\ldots$ & 0.103 & -1.95 & S09 & 1.10 & 1.10 & 0.92 & 0.35 \\
\enddata
\tablenotetext{a}{For features with hyperfine and/or isotopic structure, the value given is the total log(gf) for all lines}
\tablenotetext{b}{Sources: WB88 = \citet{WB88}, S09 = \citet{Sneden09}, DH03 = \citet{DenHartog03}, M75 = \citet{Meggers75}, L01 = \citet{Lawler01}, DH06 = \citet{DenHartog06}}
\tablenotetext{c}{Relative to literature solar values from \citet{AG89}}
\tablenotetext{d}{\citet{AG89} solar abundances are assumed for this line (see text)}
\end{deluxetable}

\subsection{Europium}  \label{subsec:Eu}

\indent Europium is an almost pure r-process element, with 98\% of solar Eu originating from the r-process. It has two main isotopes, Eu$^{151}$ and Eu$^{153}$, which make up 47.8 and 52.2\% of solar europium, respectively \citep{AG89}. Europium lines also display considerable hyperfine structure, necessitating measurement via spectral synthesis. We use the \citet{Lawler01} linelists with hyperfine and isotopic structure for the four Eu II lines with abundances presented here: 4129$\mathrm{\AA}$, 4205$\mathrm{\AA}$, 6437$\mathrm{\AA}$, and 6645$\mathrm{\AA}$. Because the KPNO, CTIO, HET, and VLT spectra have a limited wavelength range, and some of the Keck and APO spectra have low S/N in the blue, the 4129 and 4205$\mathrm{\AA}$ lines were only measured in a subset of the sample stars. Line-by-line abundances for the neutron-capture elements are given in Table \ref{tab:line_abun}; stellar average abundances are given in Table \ref{tab:stellar_avg}, as well as the stellar dispersions based on the standard deviations of the abundances given by individual lines.
\\
\indent All four of the europium lines are affected by blending to some extent. We have measured the maximum impact of blends to each line by finding the difference in measured abundances with the blend present and removed from the linelist for the Sun and in Arcturus. The 6645$\mathrm{\AA}$ line is the least blended; \citet{Lawler01} note that small Cr and Si lines are present in the solar spectrum, but these make $<$ 0.05 dex difference in the measured solar and Arcturus abundances, and CNO lines are insignificant in this wavelength region. The 6437$\mathrm{\AA}$ line has more significant blending from CNO and Si lines, particularly the Si I line at 6437.71$\mathrm{\AA}$: this line makes less than 0.10 dex difference to the Arcturus line abundance but makes up the majority of the solar profile. We adopt the Arcturus-based log(gf) for this Si line from \citet{JF13}. Eu abundances based on this line do not differ systematically from the 6645$\mathrm{\AA}$ line ($<6645 - 6437>$ = +0.02 $\pm$ 0.08 dex) so we see no reason to exclude it in our final abundances. As we discuss in Section \ref{sec:discussion}, the use of the Eu 6645 line alone instead of the stellar averages in Table \ref{tab:stellar_avg} has a minimal impact on our conclusions.
\\
\indent The 4129$\mathrm{\AA}$ and 4205$\mathrm{\AA}$ lines are larger than the lines in the red but are thought to have more severe blending, particularly from C$_2$ and CN lines. Removing these blends makes about 0.10 dex difference to the final Arcturus abundance and up to 0.25 dex difference to the solar abundance for each line. Including these large blends results in a poor fit to the solar spectrum and a low solar Eu abundance. We therefore removed the CN and C$_2$ blends from our linelists for synthesis of these two lines. The 4205$\mathrm{\AA}$ line also has a significant V blend.
\\
\indent We find an average $\epsilon_{\odot}$(Eu) of 0.52 $\pm$ 0.03 dex and an Arcturus [Eu/Fe] of 0.30 $\pm$ 0.05 dex (normalized to literature solar abundances). Our solar abundance matches the \citet{AG89} value (0.51) and the \citet{Lawler01} value (0.52) very well. A [Eu/Fe] ratio of +0.30 puts Arcturus within the range of thick disk Eu abundances for [Fe/H] $\sim$ -0.50 from \citet{Reddy06}. When normalized to our solar abundance scale, the Arcturus [Eu/Fe] = 0.29 $\pm$ 0.01; our solar and Arcturus abundances for the individual lines scale well.

\floattable
\begin{deluxetable}{l r r r r r r r r r r r r r}
\tabletypesize{\scriptsize}
\tablewidth{0pt}
\tablecolumns{14}
\tablecaption{Line-by-Line Abundances\tablenotemark{a} \label{tab:line_abun}}
\tablehead{\colhead{Cluster} & \colhead{Star} & \colhead{4129} & \colhead{4205} & \colhead{6437} & \colhead{6645} & \colhead{4085} & \colhead{4191} & \colhead{4316} & \colhead{4483} & \colhead{4498} \\ \colhead{} & \colhead{}  & \colhead{[Eu/Fe]} & \colhead{[Eu/Fe]} & \colhead{[Eu/Fe]} & \colhead{[Eu/Fe]} & \colhead{[Gd/Fe]} & \colhead{[Gd/Fe]} & \colhead{[Gd/Fe]} & \colhead{[Gd/Fe]} & \colhead{[Gd/Fe]}}
\startdata
Be17 & 265 & $\ldots$ & $\ldots$ & 0.04 & 0.10 & $\ldots$ & $\ldots$ & $\ldots$ & $\ldots$ & $\ldots$ \\
Be17 & 569 & $\ldots$ & $\ldots$ & -0.13 & -0.08 & $\ldots$ & $\ldots$ & $\ldots$ & $\ldots$ & $\ldots$ \\
Be17 & 1035 & $\ldots$ & $\ldots$ & -0.02 & 0.03 & $\ldots$ & $\ldots$ & $\ldots$ & $\ldots$ & $\ldots$ \\
Be18 & 1163 & -0.02 & -0.04 & 0.18 & 0.21 & 0.08 & 0.19 & $\ldots$ & 0.23 & 0.37 \\
Be18 & 1383 & 0.10 & 0.13 & $\ldots$ & 0.20 & 0.20 & 0.17 & 0.22 & 0.30 & 0.26 \\
Be21 & 50 & -0.02 & -0.06 & 0.04 & 0.02 & $\ldots$ & 0.37 & 0.33 & $\ldots$ & 0.26 \\
Be21 & 51 & $\ldots$ & $\ldots$ & 0.08 & 0.09 & $\ldots$ & 0.28 & 0.24 & 0.2 & 0.31 \\
Be22 & 414 & $\ldots$ & $\ldots$ & 0.3 & 0.27 & $\ldots$ & 0.18 & 0.34 & $\ldots$ & 0.28 \\
Be22 & 643 & $\ldots$ & $\ldots$ & 0.29 & 0.18 & $\ldots$ & 0.23 & 0.09 & 0.23 & 0.23 \\
\enddata
\tablenotetext{a}{Table \ref{tab:line_abun} is published in its entirety in the electronic edition of the AJ. A portion is shown here for guidance regarding its form and content.}
\end{deluxetable}

\subsection{Gadolinium}  \label{subsec:Gd}

\begin{figure*}
\epsscale{0.8}
\plotone{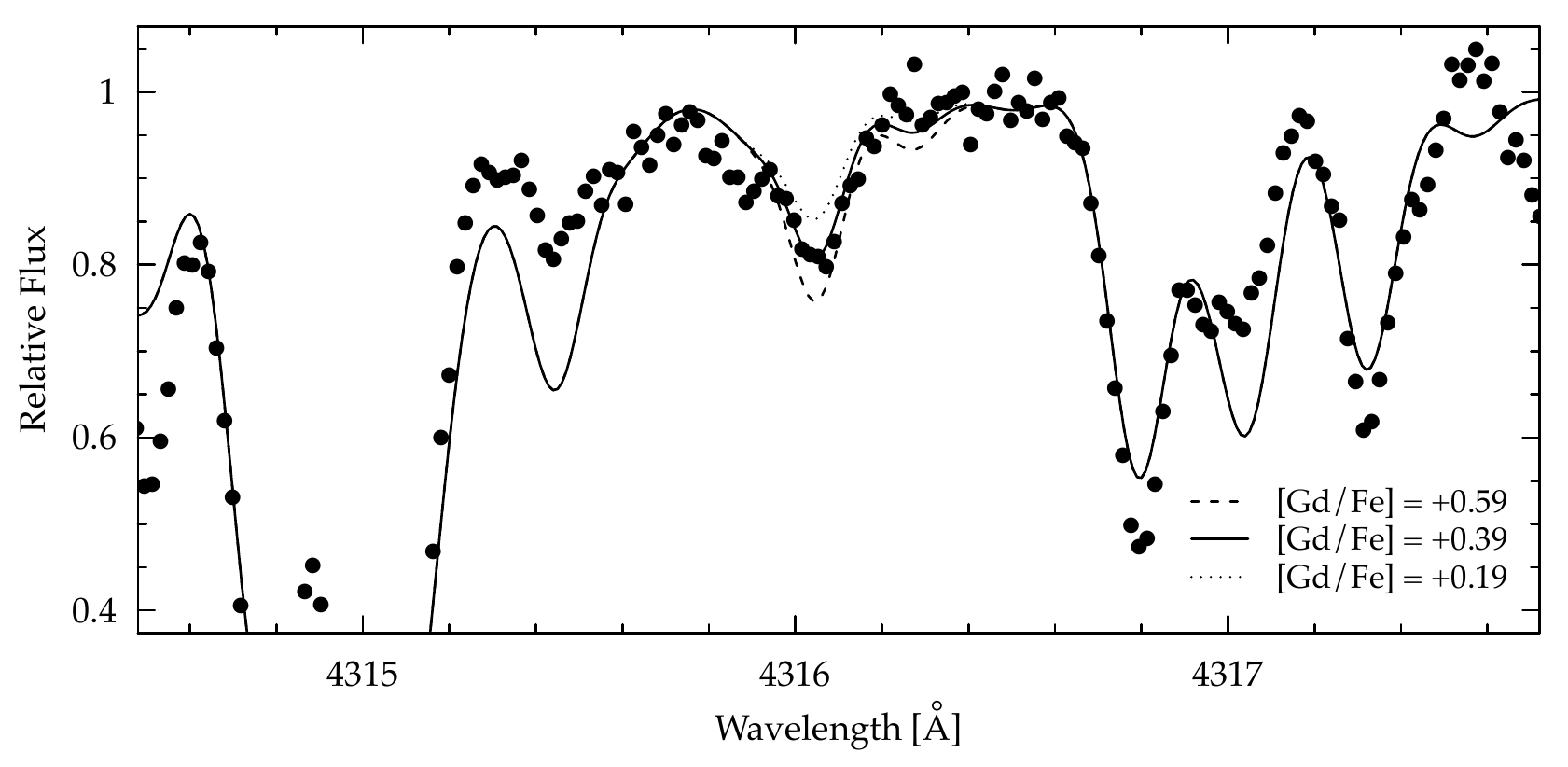}
\caption{Example fits to the 4316$\mathrm{\AA}$ gadolinium line in Be 32 star 18. Dots are the observed spectrum, the solid line represents the selected best fit, and the dashed and dotted lines are the best fit abundance +/- 0.2 dex, respectively.}
\label{Gd_fit}
\end{figure*}

\indent Gadolinium is another primarily r-process element, with 82\% of solar gadolinium originating in the r-process \citep{Sneden08}. It is comprised of seven main isotopes, Gd$^{152}$, Gd$^{154}$, Gd$^{155}$, Gd$^{156}$, Gd$^{157}$, Gd$^{158}$, and Gd$^{160}$, making up 0.3, 2.7, 14.2, 20.7, 15.7, 25.1, and 21.3\% of solar Gd, respectively. The hyperfine and isotopic structure of gadolinium have not yet been sufficiently resolved to be taken into account here, so we use single-line atomic data from \citet{DenHartog06}.
\\
\indent Unfortunately, there are no strong gadolinium lines within the wavelength range of our KPNO spectra, so we were only able to measure Gd abundances in the Keck, APO, and McDonald 2.7m spectra. Some of the lines in the solar spectrum measured by \citet{DenHartog06} are also impossible to measure given the resolution and signal-to-noise of our data, but we have isolated a set of five lines that we deem trustworthy in our spectra. The 4085, 4191, 4316, 4483, and 4498$\mathrm{\AA}$ Gd II lines are strong and clean enough to give reasonable abundances; the median dispersion on these lines for a single star is 0.08 dex. 
\\
\indent There are two lines, 4316 and 4498$\mathrm{\AA}$, that have unaccounted for blending in the solar spectrum, but since the measured Arcturus abundances for all Gd lines are similar ($\epsilon \sim 0.40$), we have included these lines assuming solar abundances from \citet{AG89}. In Figure \ref{Gd_fit} we show a sample gadolinium synthesis fit for the 4316$\mathrm{\AA}$ line for Be 32 star 18. The other three lines included have solar abundances very similar to \citet{AG89} $\epsilon$ = 1.12. Our solar measurement for the 4191$\mathrm{\AA}$ line is 0.12 dex higher than \citet{DenHartog06}, but this difference may be due to continuum placement as the line is weak in the solar spectrum. Our solar abundance based on all five lines is $\epsilon_{\odot}$ = 1.14 $\pm$ 0.02 dex, and our Arcturus [Gd/Fe] = 0.40 $\pm$ 0.05 dex.
\\
\indent Again, line-by-line abundances for the for the 20 stars in 9 clusters with blue-extended spectra are in Table \ref{tab:line_abun}, and stellar average abundances and dispersions based on the standard deviations of line abundances are given in Table \ref{tab:stellar_avg}. Due to the wavelength region they occupy, these Gd lines are blended with CN/C$_2$ and Fe features. The strength of the CN blending ranges from a maximum 0.15 dex increase in the measured Arcturus abundance in the 4085$\mathrm{\AA}$ line to no discernible effect on the 4316$\mathrm{\AA}$ line. All lines except for 4498$\mathrm{\AA}$ also appear to have some Fe blending.

\begin{figure*}
\epsscale{0.8}
\plotone{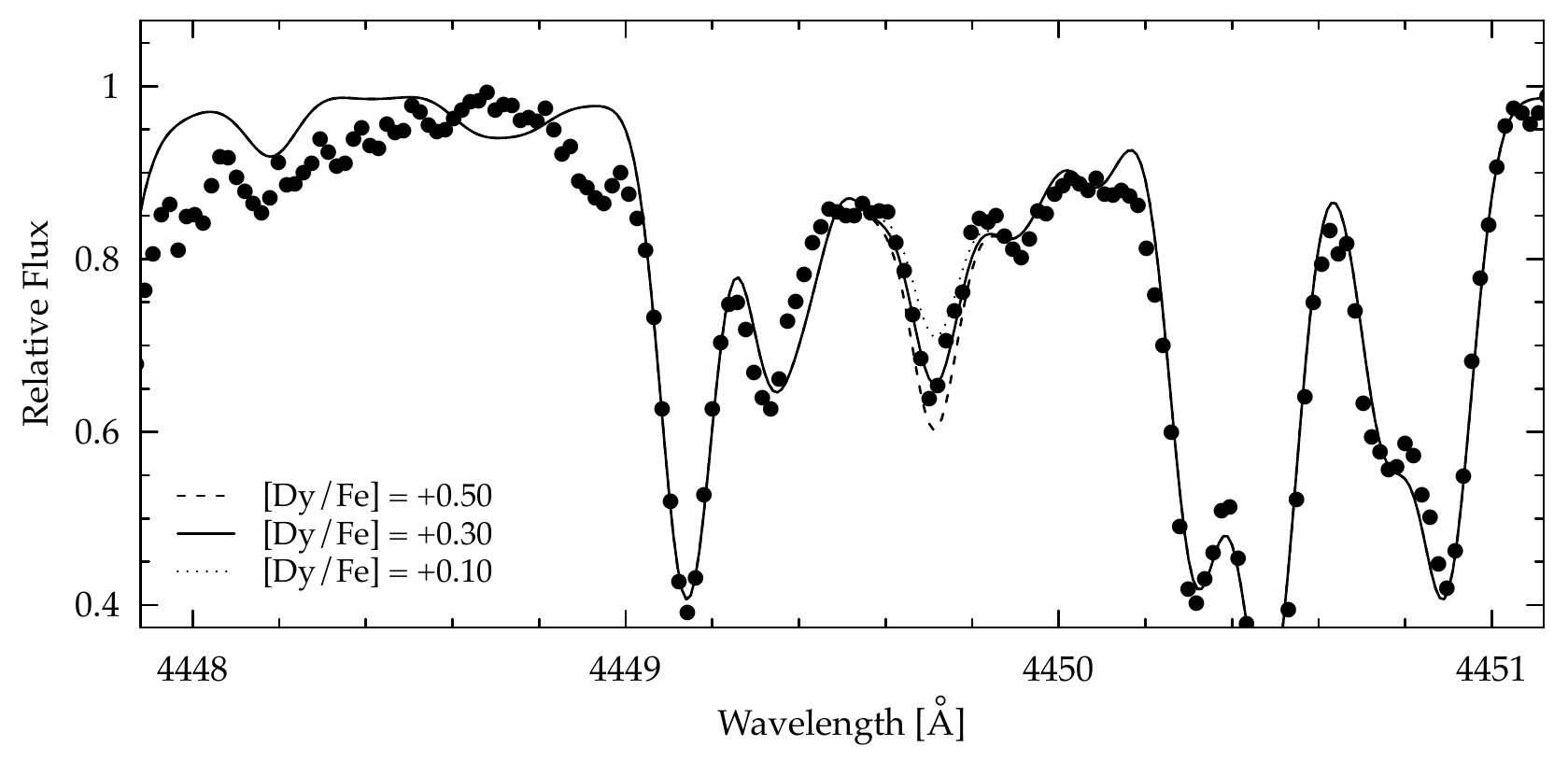}
\caption{Example fits to the 4449$\mathrm{\AA}$ dysprosium line in Be 32 star 18. Dots are the observed spectrum, the solid line represents the selected best fit, and the dashed and dotted lines are the best fit abundance +/- 0.2 dex, respectively.}
\label{Dy_fit}
\end{figure*}

\subsection{Dysprosium}  \label{subsec:Dy}

\indent Dysprosium is another primarily r-process element, with 88\% of solar dysprosium originating in the r-process. It has five primary isotopes, $^{160}$Dy, $^{161}$Dy, $^{162}$Dy, $^{163}$Dy and $^{164}$Dy, which make up 2.2\%, 19.3\%, 28.6\%, 23.2\% and 26.7\% of solar Dy, respectively \citep{Sneden08}. \citet{Sneden09} have noted that while Dy lines should have both isotopic and hyperfine structure, the contributions of these features to the line widths are small and may be ignored to fit the line as a single feature. We follow their lead in adopting the \citet{Wickliffe00} log(gf) values.
\\
\indent \citet{Sneden09} identified 8 lines in the sun that fall within the wavelength range of our Keck, APO, and McDonald 2.7m data, one of which could also be measured in the KPNO spectra. This Dy II line at 5169.69$\mathrm{\AA}$ is unfortunately weak and heavily blended, and yields consistent results only for our highest-resolution spectra. Our measurements are again limited to our high-resolution/blue-extended spectra for the strongest and least-blended Dy II lines at 4073$\mathrm{\AA}$ and 4449$\mathrm{\AA}$, and the 5169.69$\mathrm{\AA}$ line. In Figure \ref{Dy_fit} we show a sample dysprosium synthesis fit for the 4449$\mathrm{\AA}$ line selected by the program for Be 32 star 18. We have decided not to include the 4077$\mathrm{\AA}$ line, which is the most commonly measured in Dy abundance studies, because it sits on top of the Sr 4077$\mathrm{\AA}$ resonance line which is very strong in our OC stars.
\\
\indent Our solar Dy abundance based on the three lines selected is 1.12 $\pm$ 0.04, very similar to \citet{AG89} $\epsilon_{\odot}$ = 1.10. The individual solar abundances for the first two lines match \citet{Sneden09}, but they were not able to measure the 5169$\mathrm{\AA}$ line in the sun, possibly due to issues with blending. Our Arcturus [Dy/Fe] = 0.39 $\pm$ 0.03, an enhancement similar to the [Eu/Fe] and [Gd/Fe] abundances.
\\
\indent The 4073$\mathrm{\AA}$ line has significant CN blending that affects the measured Arcturus abundance by up to 0.15 dex. The 4449$\mathrm{\AA}$ line is blended with a Pr feature at 4449.83$\mathrm{\AA}$ for which hyperfine structure data is included in \citet{Sneden09} (see Section \ref{subsec:Pr}). The 5169$\mathrm{\AA}$ line has a Ti blend listed at 5169.701$\mathrm{\AA}$ in our linelist source, but including the Ti blend gives discrepant solar and Arcturus abundances so we do not include it in our final linelist. The median dispersion on our stellar Dy abundances is 0.07 dex.

\subsection{Molybdenum  \label{subsec:Mo}}

\indent Molybdenum is a mixed neutron-capture element; 32\% of solar Mo was formed via the r-process. It has five primary isotopes, $^{95}$Mo, $^{96}$Mo, $^{97}$Mo, $^{98}$Mo, and $^{100}$Mo, which are 20.4, 24.1, 12.3, 30.8, and 12.3\% of total solar Mo, respectively \citep{Sneden08}. We only have data for Mo I lines, which are weak and probably do not require hyperfine or isotopic structure. Data for structure constants of Mo are unavailable at this point; there is very little in the literature on the measurement of Mo in stars. We identify trustworthy lines from the GES linelist \citep{Heiter15}, which takes Mo atomic data from \citet{WB88}, and model each feature as a single line.
\\
\indent There are six Mo I lines in the GES linelist that are within the wavelength range of our full data set. Three of these (5506.49, 5533.03, and 5858.27$\mathrm{\AA}$) are quite weak lines with large Fe blends; the other three (5570.44, 5751.41, and 6030.64$\mathrm{\AA}$) are stronger and clean enough to measure in our OC spectra. The three lines selected have little CN blending ($<$ 0.05 dex change in the Arcturus abundance) and no major blended features. 
\\
\indent Two of the three Mo lines, 5570 and 5751$\mathrm{\AA}$, are too weak to be measured in our solar spectrum, but measured Arcturus abundances are similar, giving [Mo/Fe]$_{\mathrm{Arc.}}$ = 0.16 $\pm$ 0.03 dex, so we assume these lines have solar abundances similar to the 6030$\mathrm{\AA}$ line which matches the \citet{AG89} solar abundance.

\subsection{Praseodymium}  \label{subsec:Pr}

\indent Solar praseodymium is mostly composed of a single isotope, $^{141}$Pr, that is 50.9\% r-process. Pr lines display broad hyperfine structure that must be accounted for; we use the linelists of \citet{Sneden09} to model these effects. Several of the lines they measure in the solar spectrum are too weak or blended to measure in our spectra, but we have selected a set of four Pr lines that give consistent results for our sample stars. Be 31 star 886 and NGC 2141 star 1348 are the only spectra limited to wavelengths redward of all four lines, so they have no Pr abundances.
\\
\indent The 5219, 5259, 5322, and 5352$\mathrm{\AA}$ lines measured here have very little CNO blending. The 5219$\mathrm{\AA}$ line has a significant Co blend that composes the majority of the solar feature profile. The 5322$\mathrm{\AA}$ and 5352$\mathrm{\AA}$ lines have small Fe blends that make very little difference to measured Arcturus abundances but could affect solar abundances by up to 0.10 dex. Abundances from the 5219$\mathrm{\AA}$ line are not systematically different from the other three less blended lines, and the typical stellar dispersion between Pr line abundances is 0.07 dex, so there is no sign that the blend is causing problems with the measurement of the line. We have assumed [Co/Fe] = 0.00 for all OC stars, as most thin disk stars have approximately solar [Co/Fe] ratios \citep{Reddy03}.
\\
\indent Our Pr solar measurements are somewhat scattered; our average solar abundance is $\epsilon$ = 0.71 $\pm$ 0.10. This matches the \citet{AG89} solar Pr abundance, although our 5259 and 5322$\mathrm{\AA}$ abundances are lower than those given in \citet{Sneden09}. Different log(gf) values used for blends are probably the cause of this discrepancy. The average Arcturus [Pr/Fe] = 0.01 $\pm$ 0.03 dex.

\subsection{Neodymium}  \label{subsec:Nd}

\indent Solar Nd is comprised of seven different isotopes, $^{142}$Nd, $^{143}$Nd, $^{144}$Nd, $^{145}$Nd, $^{146}$Nd, $^{148}$Nd, and $^{150}$Nd, which are 27.2, 12.2, 23.8, 8.3, 17.2, 5.7 and 5.6\% of the total abundance, respectively \citep{Sneden08}. Approximately 42.1\% of solar Nd was created via the r-process, so we classify it as a mixed r- and s- element. Nd II has many observable lines in optical wavelengths that display minimal hyperfine and isotopic structure with $\Delta\lambda$ $<$ 0.005$\mathrm{\AA}$ \citep{DenHartog03}. Because isotopic and hyperfine structure data for these features are not generally available, and the resolution of our spectra are significantly lower than the expected wavelength deviations, we do not model this structure with synthesis.
\\
\indent We have selected a set of ten Nd II lines that appear relatively unblended in our spectra. Eight of these lines, at 5132.33, 5306.46, 5356.96, 5431.52, 5485.70, 5533.82, 5740.86, and 5811.57$\mathrm{\AA}$ have atomic data listed in \citet{DenHartog03} that are used in the GES linelist. Two of the lines, at 5416.37 and 6385.15$\mathrm{\AA}$, do not appear in \citet{DenHartog03} but are included in the GES linelist with atomic data taken from \citet{Meggers75}. The 5356.96$\mathrm{\AA}$ line is partially blended with a Sc II line at 5357.20$\mathrm{\AA}$ at the resolution of our KPNO 4m spectra, but the features are separated enough to identify that a solar [Sc/Fe] ratio fits the Sc line well for most stars, so we include these measurements in our final data set. The 5533.82$\mathrm{\AA}$ line has a significant V I blend at the same wavelength, but assuming solar [V/Fe] gives abundances in good agreement with the other lines. The median dispersion of the Nd stellar abundances is 0.08 dex.
\\
\indent We find that Nd $\epsilon_{\odot}$ = 1.49 $\pm$ 0.08; this value falls between the \citet{AG89} $\epsilon_{\odot}$ = 1.50 and \citet{DenHartog03} $\epsilon_{\odot}$ = 1.46. Our average Arcturus [Nd/Fe] = -0.03 $\pm$ 0.09 dex, but normalizing to our solar Nd abundances instead of the literature decreases the dispersion to 0.05 dex; our solar and Arcturus line abundances mostly track each other except for $\lambda$5533.

\section{Discrepant Stars}  \label{sec:discr}

\indent We have a few stars in our sample with abundances far from the cluster averages for some or all elements. Calculating cluster abundances including these stars would greatly enhance the cluster dispersions, so in some cases we have excluded them (if we have enough stars in the cluster to determine the outlier). We discuss them here.
\\
\indent NGC 2355: This cluster has a relatively high dispersion for all of the elements, including iron; the cluster [Fe/H] dispersion is 0.10 dex, while the median for our cluster sample is 0.03 dex. With only three stars, it is not completely clear which of the three, if any, is distorting the cluster average. Star 144 is +0.20 dex enhanced from the other two in Mo and Nd, but less separated in Pr and Eu. We will analyze more stars in this cluster in our forthcoming s-process focused paper, and will be able to better define the dispersions for this cluster.
\\
\indent NGC 2141: We have four stars for this cluster, two of which are newly presented in this work. Three of these stars cluster nicely, with dispersions $<$ 0.05 dex for all elements except Pr, but star 1348 has abundances $\sim$0.40 dex larger than the others for Mo and Nd (we could not measure Pr in this star due to wavelength coverage restrictions). \citet{JF13} also note that this star has enhanced Zr, Ba, and La compared to star 1007, although its Eu abundances from \citet{JF13} and this paper are not significantly higher than the other cluster members; the additional two stars (514 and 1821) allow us to confirm that 1348 is the anomalous member. We exclude this star when calculating all cluster abundances and dispersions; its enhancement in Mo and Nd, and Eu abundance within 1$\sigma$ of the cluster average, suggest that this is an s-process enhanced star.
\\
\indent NGC 6939: We have six stars for this cluster, including two from our most recent APO run. Star 190 gives enhanced abundances for all of the elements we present here, sometimes by as much as 0.5 dex. Star 121 has slightly lower abundances than the other four, though the difference is not as significant. We have excluded 190 from the final cluster abundance calculations as we have multiple stars for this cluster with more reliable abundances.

\section{Errors}  \label{sec:errors}

\indent Errors in neutron capture abundances due to errors in atmospheric and continuum setting parameters for six stars from our sample are given in Table \ref{tab:errors}. These stars encompass the full range of temperatures, gravities, microturbulent velocities, and metallicities of our sample. The values for each element were calculated by averaging the changes in all lines that were measured for each star. The changes in abundance due to changes in the continuum setting parameter are based on a two percent shift in the chosen continuum for each star. Error due to the signal-to-noise of each spectrum and the atomic data used are not contained in these values.
\\
\indent Unfortunately, our Mo lines appear to be relatively sensitive to temperature, changing abundances by $\sim$0.15 dex per 100K for all but the coolest star (N6939 31). As this is an element that has few stellar abundance measurements in the literature, we must interpret abundances for this element with caution. Dysprosium also appears to be sensitive to temperature; the Dy lines blueward of 5000$\mathrm{\AA}$ are fairly strong, but somewhat blended. Abundance errors due to changes in log(g) are fairly constant for the different stars and elements at around 0.10 dex, except for Mo which is relatively unaffected by changes in log(g). Changes in the microturbulent velocity have little effect on most stars; Dy experiences the largest effects, which are still less than 0.1 dex. The atmospheric parameter errors added in quadrature are $\sim$0.10 dex in Nd, Pr, and Eu; 0.15 dex in Mo and Gd; and 0.20 dex in Dy.
\\
\indent As mentioned in Section \ref{sec:params}, we adopt errors of $\pm$100K, 0.2 dex, and 0.2 km~s$^{-1}$ in effective temperature, log(g), and microturbulent velocity, respectively, to account for both internal errors and differences in methods from study to study. If we were to consider only internal errors of $\pm$75 K, 0.1 dex, and 0.1 km~s$^{-1}$, we would be looking at negligible errors in abundance due to microturbulent velocities for most elements, and no more than 0.05 dex for Dy; abundance errors due to log(g) would also be on the order of 0.05 dex. Abundance errors due to temperature errors would remain significant, on the order of 0.1 dex for Mo and Dy. The atmospheric parameter errors added in quadrature in this case would be $\sim$0.06 dex for Pr, Nd, Eu, and Gd; 0.11 dex for Mo; and 0.14 dex for Dy. However, this moderate decrease in atmospheric parameter errors would not significantly affect conclusions presented here as the largest contributions to abundance errors are due to errors in setting the continuum during synthesis fitting.
\\
\indent Running synthesis in a semi-automated way makes the determination of continuum errors more objective than it might otherwise be, and due to the size of the lines being measured (especially Mo) we expect these errors to be large. They are 0.10-0.15 dex for the coolest stars of the six at 4000 and 4400K, and rise to 0.2 - 0.3 dex for the hottest stars in our sample at 5200 and 5300K. The choice of the continuum uncertainty does of course depend on the signal-to-noise of the individual spectra, but for a signal-to-noise per pixel of 100 the rms dispersion of the noise is about 0.01, so our chosen continuum error is about two times the noise for a typical sample star.

\begin{deluxetable*}{l r r r r r r r r r r r r r}
\tabletypesize{\scriptsize}
\tablewidth{0pt}
\tablecolumns{14}
\tablecaption{Average Stellar Abundances \label{tab:stellar_avg}}
\tablehead{\colhead{Cluster} & \colhead{Star} & \colhead{Avg.} & \colhead{$\sigma$} & \colhead{Avg.} & \colhead{$\sigma$} & \colhead{Avg.} & \colhead{$\sigma$} & \colhead{Avg.} & \colhead{$\sigma$} & \colhead{Avg.} & \colhead{$\sigma$} & \colhead{Avg.} & \colhead{$\sigma$} \\ \colhead{} & \colhead{}  & \colhead{[Eu/Fe]} & \colhead{[Eu/Fe]} & \colhead{[Gd/Fe]} & \colhead{[Gd/Fe]} & \colhead{[Dy/Fe]} & \colhead{[Dy/Fe]} & \colhead{[Mo/Fe]} & \colhead{[Mo/Fe]} & \colhead{[Pr/Fe]} & \colhead{[Pr/Fe]} & \colhead{[Nd/Fe]} & \colhead{[Nd/Fe]}}
\startdata
Be17 & 265 & 0.07 & 0.04 & $\ldots$ & $\ldots$ & $\ldots$ & $\ldots$ & 0.01 & 0.03 & -0.10 & 0.06 & 0.01 & 0.12 \\
Be17 & 569 & -0.11 & 0.04 & $\ldots$ & $\ldots$ & $\ldots$ & $\ldots$ & -0.07 & 0.07 & -0.20 & 0.11 & -0.11 & 0.06 \\
Be17 & 1035 & 0.01 & 0.04 & $\ldots$ & $\ldots$ & $\ldots$ & $\ldots$ & -0.05 & 0.07 & -0.09 & 0.07 & 0.03 & 0.11 \\
Be18 & 1163 & 0.08 & 0.13 & 0.22 & 0.12 & 0.24 & 0.04 & 0.13 & 0.08 & 0.18 & 0.02 & 0.22 & 0.07 \\
Be18 & 1383 & 0.14 & 0.05 & 0.23 & 0.05 & 0.23 & 0.03 & 0.05 & 0.05 & 0.21 & 0.09 & 0.19 & 0.07 \\
Be21 & 50 & -0.01 & 0.04 & 0.32 & 0.06 & 0.15 & 0.18 & 0.17 & 0.08 & 0.18 & 0.05 & 0.21 & 0.05 \\
Be21 & 51 & 0.09 & 0.01 & 0.26 & 0.05 & 0.19 & 0.03 & 0.09 & 0.10 & 0.19 & 0.03 & 0.19 & 0.05 \\
Be22 & 414 & 0.29 & 0.02 & 0.27 & 0.08 & 0.31 & 0.16 & 0.06 & 0.05 & 0.22 & 0.02 & 0.32 & 0.04 \\
Be22 & 643 & 0.24 & 0.08 & 0.20 & 0.07 & 0.21 & 0.06 & 0.18 & 0.06 & 0.32 & 0.17 & 0.41 & 0.10 \\
Be31 & 886 & 0.27 & 0.07 & $\ldots$ & $\ldots$ & $\ldots$ & $\ldots$ & 0.46 & 0.17 & $\ldots$ & $\ldots$ & 0.56 & 0.08 \\
Be32 & 2 & 0.00 & 0.03 & $\ldots$ & $\ldots$ & $\ldots$ & $\ldots$ & 0.01 & 0.05 & -0.11 & 0.10 & -0.03 & 0.04 \\
Be32 & 4 & 0.06 & 0.04 & $\ldots$ & $\ldots$ & $\ldots$ & $\ldots$ & -0.01 & 0.05 & -0.12 & 0.05 & -0.04 & 0.04 \\
Be32 & 16 & 0.11 & 0.03 & 0.34 & 0.08 & 0.41 & 0.06 & 0.15 & 0.01 & 0.15 & 0.03 & 0.15 & 0.05 \\
Be32 & 18 & 0.15 & 0.04 & 0.29 & 0.11 & 0.32 & 0.10 & 0.11 & 0.01 & 0.11 & 0.03 & 0.16 & 0.05 \\
Be39 & 3 & -0.02 & 0.02 & $\ldots$ & $\ldots$ & $\ldots$ & $\ldots$ & 0.01 & 0.02 & -0.01 & 0.10 & 0.04 & 0.07 \\
Be39 & 5 & -0.01 & 0.09 & $\ldots$ & $\ldots$ & $\ldots$ & $\ldots$ & 0.07 & 0.06 & -0.01 & 0.11 & 0.05 & 0.06 \\
Be39 & 12 & 0.08 & 0.00 & $\ldots$ & $\ldots$ & $\ldots$ & $\ldots$ & 0.20 & 0.05 & 0.14 & 0.06 & 0.03 & 0.09 \\
Be39 & 14 & 0.10 & 0.06 & $\ldots$ & $\ldots$ & $\ldots$ & $\ldots$ & 0.16 & 0.01 & 0.14 & 0.08 & 0.11 & 0.08 \\
Cr261 & 1045 & -0.12 & 0.06 & $\ldots$ & $\ldots$ & $\ldots$ & $\ldots$ & -0.08 & 0.07 & -0.30 & $\ldots$ & -0.10 & 0.05 \\
Cr261 & 1080 & -0.03 & 0.01 & $\ldots$ & $\ldots$ & $\ldots$ & $\ldots$ & -0.02 & 0.07 & -0.29 & $\ldots$ & -0.10 & 0.05 \\
M67 & 105 & -0.10 & 0.03 & $\ldots$ & $\ldots$ & $\ldots$ & $\ldots$ & 0.07 & 0.03 & -0.06 & 0.03 & 0.02 & 0.04 \\
M67 & 141 & -0.07 & 0.09 & $\ldots$ & $\ldots$ & $\ldots$ & $\ldots$ & 0.07 & 0.04 & -0.05 & 0.05 & -0.01 & 0.12 \\
M67 & 170 & -0.15 & 0.04 & $\ldots$ & $\ldots$ & $\ldots$ & $\ldots$ & 0.01 & 0.07 & -0.07 & 0.04 & -0.01 & 0.05 \\
N188 & 532 & -0.03 & 0.06 & $\ldots$ & $\ldots$ & $\ldots$ & $\ldots$ & 0.38 & 0.08 & 0.00 & 0.12 & 0.19 & 0.06 \\
N188 & 747 & -0.11 & 0.01 & $\ldots$ & $\ldots$ & $\ldots$ & $\ldots$ & 0.17 & 0.06 & -0.08 & 0.06 & 0.08 & 0.11 \\
N188 & 919 & -0.19 & 0.13 & $\ldots$ & $\ldots$ & $\ldots$ & $\ldots$ & 0.15 & 0.13 & -0.07 & 0.10 & 0.11 & 0.07 \\
N188 & 1224 & -0.15 & 0.05 & $\ldots$ & $\ldots$ & $\ldots$ & $\ldots$ & 0.23 & 0.04 & -0.03 & 0.04 & 0.17 & 0.08 \\
N1193 & 282 & -0.03 & 0.09 & $\ldots$ & $\ldots$ & $\ldots$ & $\ldots$ & 0.28 & 0.08 & 0.14 & 0.10 & 0.05 & 0.07 \\
N1245 & 10 & -0.26 & 0.02 & $\ldots$ & $\ldots$ & $\ldots$ & $\ldots$ & 0.21 & $\ldots$ & -0.11 & 0.05 & 0.01 & 0.12 \\
N1245 & 125 & -0.09 & 0.05 & $\ldots$ & $\ldots$ & $\ldots$ & $\ldots$ & 0.02 & 0.06 & 0.09 & 0.11 & 0.19 & 0.10 \\
N1245 & 160 & -0.06 & 0.05 & $\ldots$ & $\ldots$ & $\ldots$ & $\ldots$ & 0.09 & $\ldots$ & 0.15 & 0.07 & 0.06 & 0.09 \\
N1245 & 382 & -0.10 & 0.01 & $\ldots$ & $\ldots$ & $\ldots$ & $\ldots$ & 0.15 & $\ldots$ & 0.13 & 0.05 & 0.08 & 0.05 \\
N1817 & 73 & -0.04 & 0.00 & $\ldots$ & $\ldots$ & $\ldots$ & $\ldots$ & 0.34 & 0.01 & 0.02 & 0.09 & 0.16 & 0.07 \\
N1817 & 79 & -0.12 & 0.01 & $\ldots$ & $\ldots$ & $\ldots$ & $\ldots$ & 0.22 & 0.13 & 0.13 & 0.06 & 0.14 & 0.06 \\
N1817 & 206 & -0.11 & 0.03 & 0.05 & 0.15 & -0.02 & 0.11 & 0.27 & $\ldots$ & 0.02 & 0.08 & 0.14 & 0.08 \\
N1817 & 1456 & -0.07 & 0.06 & 0.03 & 0.22 & -0.02 & 0.14 & 0.17 & 0.06 & 0.13 & 0.09 & 0.15 & 0.07 \\
N1883 & 8 & -0.22 & 0.02 & $\ldots$ & $\ldots$ & $\ldots$ & $\ldots$ & -0.19 & 0.05 & 0.05 & 0.06 & 0.00 & 0.08 \\
N1883 & 9 & -0.18 & 0.02 & $\ldots$ & $\ldots$ & $\ldots$ & $\ldots$ & 0.11 & 0.16 & 0.03 & 0.12 & 0.03 & 0.10 \\
N2141 & 514 & -0.16 & 0.01 & 0.02 & 0.10 & $\ldots$ & $\ldots$ & -0.07 & 0.12 & -0.10 & 0.12 & -0.10 & 0.09 \\
N2141 & 1007 & -0.12 & 0.04 & $\ldots$ & $\ldots$ & $\ldots$ & $\ldots$ & -0.08 & 0.01 & -0.05 & 0.06 & -0.02 & 0.08 \\
N2141 & 1348 & -0.11 & 0.05 & $\ldots$ & $\ldots$ & $\ldots$ & $\ldots$ & 0.27 & 0.06 & $\ldots$ & $\ldots$ & 0.25 & 0.04 \\
N2141 & 1821 & -0.21 & $\ldots$ & 0.06 & 0.10 & $\ldots$ & $\ldots$ & -0.13 & 0.10 & -0.29 & 0.10 & -0.08 & 0.10 \\
N2158 & 4230 & -0.23 & 0.08 & $\ldots$ & $\ldots$ & $\ldots$ & $\ldots$ & -0.02 & 0.04 & 0.00 & 0.06 & 0.06 & 0.08 \\
N2194 & 55 & -0.16 & 0.11 & $\ldots$ & $\ldots$ & $\ldots$ & $\ldots$ & 0.21 & 0.11 & 0.10 & 0.06 & 0.25 & 0.09 \\
N2194 & 57 & -0.29 & 0.03 & $\ldots$ & $\ldots$ & $\ldots$ & $\ldots$ & 0.29 & 0.12 & 0.13 & 0.09 & 0.13 & 0.10 \\
N2355 & 144 & 0.02 & 0.05 & $\ldots$ & $\ldots$ & $\ldots$ & $\ldots$ & 0.41 & $\ldots$ & 0.33 & $\ldots$ & 0.36 & 0.08 \\
N2355 & 398 & -0.09 & 0.07 & $\ldots$ & $\ldots$ & $\ldots$ & $\ldots$ & 0.24 & 0.08 & 0.08 & 0.04 & 0.10 & 0.09 \\
N2355 & 668 & -0.12 & 0.04 & $\ldots$ & $\ldots$ & $\ldots$ & $\ldots$ & 0.23 & 0.04 & 0.14 & 0.06 & 0.16 & 0.09 \\
N6192 & 9 & -0.19 & 0.12 & $\ldots$ & $\ldots$ & $\ldots$ & $\ldots$ & -0.06 & 0.02 & -0.14 & 0.06 & -0.10 & 0.05 \\
N6192 & 45 & -0.05 & 0.03 & $\ldots$ & $\ldots$ & $\ldots$ & $\ldots$ & 0.18 & 0.06 & 0.10 & 0.06 & 0.15 & 0.06 \\
N6192 & 96 & -0.15 & 0.05 & $\ldots$ & $\ldots$ & $\ldots$ & $\ldots$ & 0.06 & 0.01 & -0.07 & 0.03 & 0.03 & 0.03 \\
N6192 & 137 & -0.06 & 0.03 & $\ldots$ & $\ldots$ & $\ldots$ & $\ldots$ & 0.02 & 0.01 & -0.03 & 0.06 & 0.08 & 0.06 \\
N6939 & 31 & -0.21 & 0.05 & $\ldots$ & $\ldots$ & $\ldots$ & $\ldots$ & -0.19 & 0.03 & -0.16 & 0.04 & -0.08 & 0.06 \\
N6939 & 53 & -0.31 & 0.05 & -0.01 & 0.15 & -0.10 & 0.12 & 0.10 & 0.07 & 0.02 & 0.11 & 0.01 & 0.10 \\
N6939 & 65 & -0.37 & 0.07 & -0.17 & 0.17 & -0.12 & 0.12 & 0.00 & 0.04 & 0.08 & 0.08 & -0.04 & 0.05 \\
N6939 & 121 & -0.28 & 0.01 & $\ldots$ & $\ldots$ & $\ldots$ & $\ldots$ & -0.38 & 0.06 & -0.27 & 0.08 & -0.26 & 0.07 \\
N6939 & 190 & -0.04 & $\ldots$ & $\ldots$ & $\ldots$ & $\ldots$ & $\ldots$ & 0.41 & 0.08 & 0.33 & 0.08 & 0.49 & 0.05 \\
N6939 & 212 & -0.10 & 0.13 & $\ldots$ & $\ldots$ & $\ldots$ & $\ldots$ & -0.01 & 0.13 & -0.05 & 0.08 & 0.03 & 0.08 \\
N7142 & 196 & -0.12 & 0.11 & $\ldots$ & $\ldots$ & $\ldots$ & $\ldots$ & 0.11 & 0.08 & 0.03 & 0.04 & 0.02 & 0.06 \\
N7142 & 229 & -0.15 & 0.10 & $\ldots$ & $\ldots$ & $\ldots$ & $\ldots$ & 0.15 & 0.05 & -0.06 & 0.09 & 0.01 & 0.06 \\
N7142 & 377 & -0.16 & 0.06 & $\ldots$ & $\ldots$ & $\ldots$ & $\ldots$ & 0.12 & 0.01 & -0.01 & 0.04 & -0.02 & 0.06 \\
N7142 & 421 & -0.14 & 0.06 & $\ldots$ & $\ldots$ & $\ldots$ & $\ldots$ & 0.17 & 0.06 & -0.06 & 0.08 & 0.03 & 0.07 \\
N7789 & 212 & -0.07 & 0.02 & 0.18 & 0.23 & 0.02 & $\ldots$ & 0.17 & 0.04 & 0.07 & 0.05 & 0.15 & 0.06 \\
N7789 & 468 & -0.15 & 0.08 & 0.14 & 0.00 & 0.04 & 0.04 & 0.14 & 0.03 & 0.13 & 0.10 & 0.12 & 0.10 \\
N7789 & 605 & -0.09 & 0.01 & 0.14 & 0.01 & -0.03 & $\ldots$ & 0.20 & 0.18 & 0.05 & 0.06 & 0.17 & 0.09 \\
N7789 & 765 & -0.14 & 0.15 & 0.19 & 0.01 & 0.03 & 0.07 & 0.18 & 0.02 & 0.17 & 0.10 & 0.17 & 0.05 \\
N7789 & 958 & -0.10 & $\ldots$ & 0.19 & 0.01 & 0.11 & 0.00 & 0.43 & $\ldots$ & 0.18 & 0.04 & 0.20 & 0.08 \\
PWM4 & RGB1 & 0.10 & 0.01 & 0.07 & 0.05 & 0.18 & $\ldots$ & -0.06 & 0.04 & 0.12 & 0.09 & 0.14 & 0.06 \\
\enddata
\end{deluxetable*}

\floattable
\begin{deluxetable}{l l r r r r}
\tabletypesize{\scriptsize}
\tablewidth{0pt}
\tablecolumns{6}
\tablecaption{Abundance Errors Due to Atmospheric Parameter Errors  \label{tab:errors}}
\tablehead{\colhead{Star} & \colhead{El.} & \colhead{T $\pm$} & \colhead{log(g) $\pm$} & \colhead{$\xi$ $\pm$} & \colhead{Cont.} \\ \colhead{} & \colhead{} & \colhead{100K} & \colhead{0.2 dex} & \colhead{0.2 km~s$^{-1}$} & \colhead{$\pm$ 2\%} }
\startdata
Be18 1383\tablenotemark{a} & $\Delta$Mo & 0.14 & 0.03 & 0.01 & 0.14 \\
 & $\Delta$Pr & 0.03 & 0.09 & 0.01 & 0.09 \\
 & $\Delta$Nd & 0.02 & 0.10 & 0.02 & 0.13 \\
 & $\Delta$Eu & 0.04 & 0.12 & 0.05 & 0.09 \\
 & $\Delta$Gd & 0.06 & 0.11 & 0.05 & 0.12 \\
 & $\Delta$Dy & 0.19 & 0.15 & 0.08 & 0.13 \\
 \tableline
Be32 16\tablenotemark{b} & $\Delta$Mo & 0.15 & 0.02 & 0.01 & 0.21 \\
 & $\Delta$Pr & 0.03 & 0.09 & 0.01 & 0.17 \\
 & $\Delta$Nd & 0.03 & 0.10 & 0.01 & 0.22 \\
 & $\Delta$Eu & 0.02 & 0.11 & 0.02 & 0.10 \\
 & $\Delta$Gd & 0.06 & 0.09 & 0.04 & 0.14 \\
 & $\Delta$Dy & 0.12 & 0.10 & 0.07 & 0.11 \\
  \tableline
N2355 144\tablenotemark{c} & $\Delta$Mo & 0.14 & 0.00 & 0.01 & 0.29 \\
 & $\Delta$Pr & 0.04 & 0.09 & 0.00 & 0.25 \\
 & $\Delta$Nd & 0.03 & 0.09 & 0.01 & 0.20 \\
 & $\Delta$Eu & 0.03 & 0.11 & 0.01 & 0.35 \\
  \tableline
N2355 398\tablenotemark{d} & $\Delta$Mo & 0.17 & 0.01 & 0.01 & 0.27 \\
 & $\Delta$Pr & 0.04 & 0.09 & 0.01 & 0.14 \\
 & $\Delta$Nd & 0.03 & 0.09 & 0.01 & 0.15 \\
 & $\Delta$Eu & 0.03 & 0.10 & 0.01 & 0.18 \\
  \tableline
N6192 9\tablenotemark{e} & $\Delta$Mo & 0.18 & 0.03 & 0.01 & 0.18 \\
 & $\Delta$Pr & 0.04 & 0.10 & 0.01 & 0.11 \\
 & $\Delta$Nd & 0.03 & 0.10 & 0.02 & 0.17 \\
 & $\Delta$Eu & 0.02 & 0.12 & 0.00 & 0.16 \\
  \tableline
N6939 31\tablenotemark{f} & $\Delta$Mo & 0.09 & 0.07 & 0.06 & 0.12 \\
 & $\Delta$Pr & 0.04 & 0.10 & 0.03 & 0.09 \\
 & $\Delta$Nd & 0.01 & 0.12 & 0.05 & 0.13 \\
 & $\Delta$Eu & 0.07 & 0.13 & 0.01 & 0.14 \\
\enddata
\tablenotetext{a}{T = 4400K, log(g) = 1.9 dex, $\xi$ = 1.30 km~s$^{-1}$, [Fe/H] = -0.34}
\tablenotetext{b}{T = 4900K, log(g) = 2.7 dex, $\xi$ = 1.00 km~s$^{-1}$, [Fe/H] = -0.26}
\tablenotetext{c}{T = 5300K, log(g) = 3.3 dex, $\xi$ = 1.30 km~s$^{-1}$, [Fe/H] = -0.14}
\tablenotetext{d}{T = 5200K, log(g) = 2.1 dex, $\xi$ = 1.90 km~s$^{-1}$, [Fe/H] = +0.05}
\tablenotetext{e}{T = 4900K, log(g) = 1.9 dex, $\xi$ = 1.60 km~s$^{-1}$, [Fe/H] = +0.17}
\tablenotetext{f}{T = 4000K, log(g) = 0.9 dex, $\xi$ = 1.50 km~s$^{-1}$, [Fe/H] = -0.02}
\end{deluxetable}

\section{Abundance Discussion and Literature Comparison}  \label{sec:discussion}

\subsection{OC Literature Comparison}  \label{subsec:lit}

\indent Cluster average abundances are given in Table \ref{tab:OC_avg}, along with dispersions (the dispersion between stars, or individual features for clusters with one star). There are a few large sets of OC neutron-capture measurements in the literature, including \citet{Yong05, Yong12}, and \citet{Reddy12, Reddy13, Reddy15} which include some overlap with our sample that allows direct comparison. There are also several analyses of smaller numbers of clusters that overlap with our sample; overall, 10 of our 23 sample clusters have Eu abundances in the literature, five of which also have Nd abundances, and one (NGC 7789) that has a Pr abundance. M67 is a particularly well-studied OC with several Eu measurements in the literature. Table \ref{tab:lit_ab_comp} shows relative cluster abundances for elements with literature OC measurements in the sense $\Delta$[X/Fe] = [X/Fe]$_{lit.}$ - [X/Fe]$_{our}$. All abundances in Table \ref{tab:lit_ab_comp} have been placed on our abundance scale (including the hyperfine structure correction for Eu abundances as discussed below) except for \citet{Taut00, Taut05} which do not list adopted solar abundances.
\\
\indent The neutron-capture elements are the focus of this work, but because [Fe/H] abundances form the baseline for our abundance measurements and because many questions still remain about the Galactic [Fe/H] gradient (we discuss this further in later sections), we must first consider systematic offsets in our iron abundances. We find that our cluster [Fe/H] values are $\sim$0.10 dex higher than those in the literature, and $\sim$0.15 dex higher than those found by \citet{Yong05, Yong12}. Differences of this magnitude from study to study are not uncommon, and fall within the uncertainty due to atmospheric parameter errors and continuum setting (we adopt uncertainties due to atmospheric parameters for [Fe/H] from \citet{JF13} since most [Fe/H] cluster measurements were taken from that work). We do not attempt to correct for this relatively small effect but we will consider it further when discussing the neutron-capture literature comparisons.
\\
\indent Because Eu lines display hyperfine and isotopic structure, an abundance comparison for these elements is not as straightforward as equivalent width measurements. \citet{Reddy12, Reddy13, Reddy15} use hyperfine and isotopic structure data from \citet{Mucciarelli08}, so we ran syntheses for some of our own clusters and found a consistent abundance increase of 0.30 dex using the Mucciarelli Eu data instead of \citet{Lawler01}, so we correct for this effect as well as the solar abundance offset. This adjustment due to the differing atomic data used for Eu moves the [Eu/Fe] abundances from \citet{Reddy12, Reddy13, Reddy15} into the same range as our young cluster averages. 
\\
\indent There is, however, a consistent offset between our [Eu/Fe] measurements and \citet{Yong05, Yong12}, which is more puzzling because they also use the \citet{Lawler01} atomic data. This may be partially explained by the systematic differences in cluster [Fe/H] abundances, but even when considering [Eu/H] our values are still $\sim$0.1 dex lower than those of \citet{Yong05, Yong12}. This difference is within the expected error due to stellar atmospheric parameters or continuum fitting. \citet{Yong05, Yong12} find higher gravities for many of the stars we have also analyzed. Their spectroscopically determined log(g)'s are an average of 0.3 dex higher than ours for the two Be18, Be21, Be22, Be32, and single PWM4 star, which, according to our determination of atmospheric parameter errors (see Table \ref{tab:errors}), is enough to increase the measured Eu abundance by 0.1 - 0.2 dex.
\\
\indent Generally, literature studies of Eu in OCs rely on measurements of the 6645$\mathrm{\AA}$ line alone because it is the least blended of the Eu lines, and we must consider possible effects of the introduction of the other three lines presented here. Our Eu cluster abundances, when calculated only with the 6645$\mathrm{\AA}$ line, never differ by more than 0.1 dex from the abundance based on all four lines, with a median difference of 0.03 dex that does not appear to be systematic. We find little difference in trends with Galactocentric radius or age (discussed in the next two sections).
\\
\indent Only one OC, NGC 3680, has a Gd measurement in the literature. \citet{Mitschang12} find a [Gd/Fe] = -0.45 $\pm$ 0.09 dex, which falls significantly below the range of our [Gd/Fe] measurements even for clusters with similar properties ($\sim$1.5 Gyr, R$_{\mathrm{GC}}$ = 8 kpc). This Gd abundance relies on a single line at 4463$\mathrm{\AA}$. \citet{DenHartog06} calculate atomic data for this line but do not report a solar abundance measurement, and we find that this line is too severely blended for measurement in our sample stars. There are literature measurements of Gd in metal-poor, r-process enriched stars, but they do not cover the same metallicity range as our OC and are more difficult to place within the context of this study, so we do not include them here. 
\\
\indent To our knowledge, no measurements of open cluster Dy abundances are present in the literature. \citet{Francois07} have a compilation of literature measurements suggesting that [Dy/Fe] abundances decrease with increasing [Fe/H] and also show less scatter, although the abundance measurements are for stars with -4 $<$ [Fe/H] $<$ -1, well outside of the typical OC metallicity range. The [Dy/Fe] abundances appear to be approaching solar at an [Fe/H] of -1. 
\\
\indent Very little observational data for Mo is available in the literature due to the difficulty of measuring weak, singly ionized Mo lines in the optical. One cluster from our sample, NGC 7789, has a Pr abundance measurement in the literature, from \citet{Taut05}. Again, the two abundance determinations agree within the errors. \citet{Taut05} measure the 5322$\mathrm{\AA}$ line using atomic data from the VALD.
\\
\indent The \citet{Reddy12, Reddy13, Reddy15} OC sample also includes Nd abundances. The Nd data in the literature is measured via equivalent widths, and as we mentioned previously Nd has minimal hyperfine and isotopic splitting, so we have simply corrected these based on the offset in solar abundance measurements for [Nd/Fe], which is negligible here. Our Nd abundances match up well with the handful available in the literature for our OCs with differences on the order of $\sim$0.05 dex which do not appear to be systematic. The \citet{Reddy12, Reddy13, Reddy15} and \citet{Pancino10} Nd abundances are based on a few lines with atomic data from \citet{DenHartog03}, and \citet{CP11} use VALD atomic data that is very similar for the three lines measured, so this agreement is reassuring but not surprising.

\floattable
\begin{deluxetable}{l r r r r r r r r r r r r r r}
\tabletypesize{\scriptsize}
\tablewidth{0pt}
\tablecolumns{15}
\tablecaption{Cluster Average Abundances  \label{tab:OC_avg}}
\tablehead{\colhead{} & \colhead{Avg.} & \colhead{} & \colhead{Avg.} & \colhead{} & \colhead{Avg.} & \colhead{} & \colhead{Avg.} & \colhead{} & \colhead{Avg.} & \colhead{} & \colhead{Avg.} & \colhead{} & \colhead{Avg.} & \colhead{} \\ \colhead{Cluster} & \colhead{[Fe/H]} & \colhead{$\sigma$Fe} & \colhead{[Eu/Fe]} & \colhead{$\sigma$Eu} & \colhead{[Gd/Fe]} & \colhead{$\sigma$Gd} & \colhead{[Dy/Fe]} & \colhead{$\sigma$Dy} & \colhead{[Mo/Fe]} & \colhead{$\sigma$Mo} & \colhead{[Pr/Fe]} & \colhead{$\sigma$Pr} & \colhead{[Nd/Fe]} & \colhead{$\sigma$Nd}}
\startdata
Be17 & -0.12 & 0.01 & -0.01 & 0.09 & $\ldots$ & $\ldots$ & $\ldots$ & $\ldots$ & -0.03 & 0.04 & -0.13 & 0.06 & -0.02 & 0.08 \\
Be18 & -0.32 & 0.03 & 0.11 & 0.04 & 0.22 & 0.01 & 0.23 & 0.01 & 0.09 & 0.05 & 0.20 & 0.02 & 0.20 & 0.03 \\
Be21 & -0.21 & 0.10 & 0.04 & 0.06 & 0.29 & 0.04 & 0.17 & 0.03 & 0.13 & 0.06 & 0.18 & 0.01 & 0.20 & 0.01 \\
Be22 & -0.27 & 0.04 & 0.26 & 0.04 & 0.23 & 0.05 & 0.26 & 0.07 & 0.12 & 0.08 & 0.27 & 0.07 & 0.36 & 0.07 \\
Be31 & -0.35 & $\ldots$ & 0.27 & 0.07 & $\ldots$ & $\ldots$ & $\ldots$ & $\ldots$ & 0.46 & 0.17 & $\ldots$ & $\ldots$ & 0.56 & 0.08 \\
Be32 & -0.26 & 0.06 & 0.08 & 0.07 & 0.32 & 0.03 & 0.37 & 0.06 & 0.06 & 0.07 & 0.01 & 0.14 & 0.06 & 0.11 \\
Be39 & -0.13 & 0.02 & 0.04 & 0.06 & $\ldots$ & $\ldots$ & $\ldots$ & $\ldots$ & 0.11 & 0.09 & 0.07 & 0.09 & 0.06 & 0.03 \\
Cr261 & -0.06 & 0.04 & -0.07 & 0.07 & $\ldots$ & $\ldots$ & $\ldots$ & $\ldots$ & -0.05 & 0.04 & -0.30 & 0.01 & -0.10 & 0.00 \\
M67 & 0.05 & 0.04 & -0.11 & 0.04 & $\ldots$ & $\ldots$ & $\ldots$ & $\ldots$ & 0.05 & 0.03 & -0.06 & 0.01 & 0.00 & 0.02 \\
N188 & 0.12 & 0.04 & -0.12 & 0.07 & $\ldots$ & $\ldots$ & $\ldots$ & $\ldots$ & 0.23 & 0.10 & -0.05 & 0.04 & 0.14 & 0.05 \\
N1193 & -0.17 & $\ldots$ & -0.03 & 0.09 & $\ldots$ & $\ldots$ & $\ldots$ & $\ldots$ & 0.28 & 0.08 & 0.14 & 0.10 & 0.05 & 0.07 \\
N1245 & 0.02 & 0.03 & -0.12 & 0.09 & $\ldots$ & $\ldots$ & $\ldots$ & $\ldots$ & 0.12 & 0.08 & 0.07 & 0.12 & 0.08 & 0.08 \\
N1817 & -0.05\tablenotemark{a} & 0.02 & -0.08 & 0.04 & 0.04 & 0.02 & -0.02 & 0.00 & 0.25 & 0.07 & 0.07 & 0.07 & 0.15 & 0.01 \\
N1883 & -0.04 & 0.01 & -0.20 & 0.03 & $\ldots$ & $\ldots$ & $\ldots$ & $\ldots$ & -0.04 & 0.21 & 0.04 & 0.02 & 0.01 & 0.02 \\
N2141 & -0.09\tablenotemark{a} & 0.01 & -0.16 & 0.05 & 0.04 & 0.03 & $\ldots$ & $\ldots$ & -0.09 & 0.03 & -0.15 & 0.13 & -0.06 & 0.04 \\
N2158 & -0.05 & $\ldots$ & -0.23 & 0.08 & $\ldots$ & $\ldots$ & $\ldots$ & $\ldots$ & -0.02 & 0.04 & 0.00 & 0.06 & 0.06 & 0.08 \\
N2194 & -0.06 & 0.00 & -0.23 & 0.09 & $\ldots$ & $\ldots$ & $\ldots$ & $\ldots$ & 0.25 & 0.05 & 0.12 & 0.02 & 0.19 & 0.08 \\
N2355 & -0.04 & 0.10 & -0.06 & 0.07 & $\ldots$ & $\ldots$ & $\ldots$ & $\ldots$ & 0.29 & 0.10 & 0.18 & 0.13 & 0.20 & 0.14 \\
N6192 & 0.10 & 0.05 & -0.11 & 0.07 & $\ldots$ & $\ldots$ & $\ldots$ & $\ldots$ & 0.05 & 0.10 & -0.03 & 0.10 & 0.04 & 0.10 \\
N6939 & 0.03\tablenotemark{a} & 0.06 & -0.25 & 0.10 & -0.09 & 0.11 & -0.11 & 0.01 & -0.10 & 0.19 & -0.08 & 0.14 & -0.07 & 0.11 \\
N7142 & 0.08 & 0.02 & -0.14 & 0.02 & $\ldots$ & $\ldots$ & $\ldots$ & $\ldots$ & 0.14 & 0.03 & -0.02 & 0.04 & 0.01 & 0.02 \\
N7789 & 0.00 & 0.03 & -0.11 & 0.03 & 0.17 & 0.03 & 0.03 & 0.03 & 0.22 & 0.12 & 0.12 & 0.06 & 0.16 & 0.03 \\
PWM4 & -0.18 & $\ldots$ & 0.10 & 0.01 & 0.07 & 0.05 & 0.18 & $\ldots$ & -0.06 & 0.04 & 0.12 & 0.09 & 0.14 & 0.06 \\
\enddata
\tablenotetext{a}{Based on stellar [Fe/H] measurements presented here as well as those in \citet{JF13}}
\end{deluxetable}

\floattable
\begin{deluxetable}{l c r r r r r r r}
\tabletypesize{\scriptsize}
\tablewidth{0pt}
\tablecolumns{9}
\tablecaption{Literature Abundance Comparisons for Clusters  \label{tab:lit_ab_comp}}
\tablehead{\colhead{Cluster} & \colhead{Source\tablenotemark{a}} & \colhead{$\Delta$[Fe/H]} & \colhead{$\Delta$[Eu/Fe]} & \colhead{$\sigma_{Eu}$\tablenotemark{b}} & \colhead{$\Delta$[Pr/Fe]} & \colhead{$\sigma_{Pr}$\tablenotemark{b}} & \colhead{$\Delta$[Nd/Fe]} & \colhead{$\sigma_{Nd}$\tablenotemark{b}}}
\startdata
Be18 & Y12 & -0.14 & 0.22 & 0.06 & $\ldots$ & $\ldots$ & $\ldots$ & $\ldots$ \\
Be21 & Y12 & -0.12 & 0.30 & 0.08 & $\ldots$ & $\ldots$ & $\ldots$ & $\ldots$ \\
Be22 & Y12 & -0.19 & 0.03 & 0.10 & $\ldots$ & $\ldots$ & $\ldots$ & $\ldots$ \\
Be31 & Y05 & -0.24 & 0.32 & $\ldots$ & $\ldots$ & $\ldots$ & $\ldots$ & $\ldots$ \\
Be32 & Y12 & -0.14 & 0.26 & 0.11 & $\ldots$ & $\ldots$ & $\ldots$ & $\ldots$ \\
Be32 & CP11 & -0.06 & $\ldots$ & $\ldots$ & $\ldots$ & $\ldots$ & -0.08 & 0.06 \\
M67 & PCRG10 & -0.02 & $\ldots$ & $\ldots$ & $\ldots$ & $\ldots$ & 0.03 & 0.05 \\
M67 & T00 & -0.09 & 0.12 & 0.07 & $\ldots$ & $\ldots$ & $\ldots$ & $\ldots$ \\
M67 & Y05 & -0.05 & 0.20 & 0.02 & $\ldots$ & $\ldots$ & $\ldots$ & $\ldots$ \\
M67 & RGL & -0.11 & -0.11 & $\ldots$ & $\ldots$ & $\ldots$ & 0.01 & 0.04 \\
N1817 & RGL & -0.06 & -0.09 & $\ldots$ & $\ldots$ & $\ldots$ & -0.02 & 0.04 \\
N2141 & Y05 & -0.11 & 0.36 & $\ldots$ & $\ldots$ & $\ldots$ & $\ldots$ & $\ldots$ \\
N7789 & PCRG10 & 0.02 & $\ldots$ & $\ldots$ & $\ldots$ & $\ldots$ & -0.04 & 0.30 \\
N7789 & T05 & -0.04 & 0.13 & 0.12 & -0.04 & 0.05 & $\ldots$ & $\ldots$ \\
PWM4 & Y12 & -0.18 & 0.05 & $\ldots$ & $\ldots$ & $\ldots$ & $\ldots$ & $\ldots$ \\
\enddata
\tablenotetext{a}{References for literature comparisons: CP11 = \citet{CP11}; PCRG10 = \citet{Pancino10}; RGL = \citet{Reddy12, Reddy13}; T00 = \citet{Taut00}; T05 = \citet{Taut05}; Y05 = \citet{Yong05}}
\tablenotetext{b}{These columns are the errors on cluster measurements given in the literature sources.}
\end{deluxetable}

\subsection{Age Trends}  \label{subsec:age}

\begin{figure*}
\epsscale{0.6}
\plotone{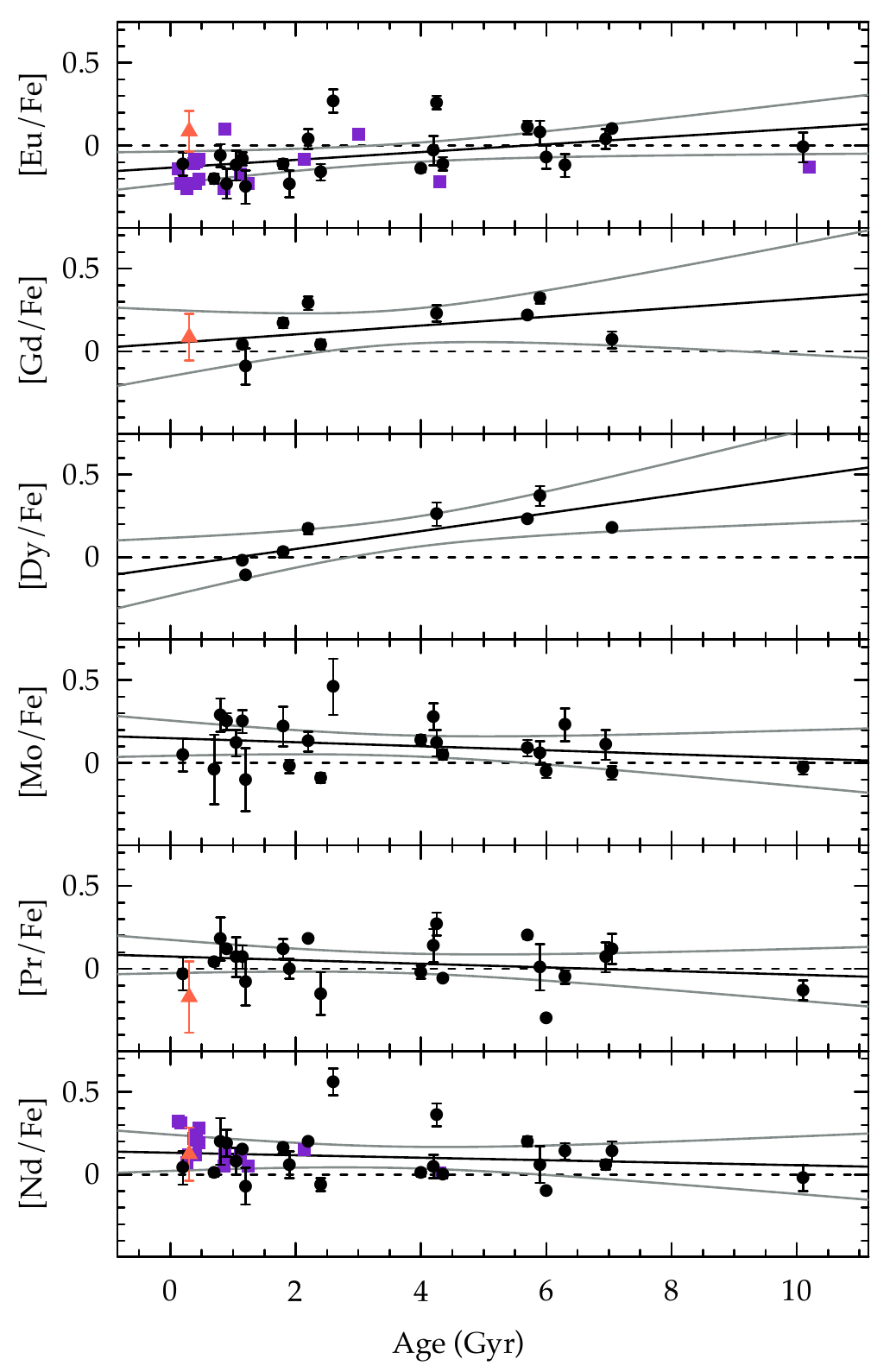}
\caption{Abundance trends with age for OCs from this paper (black with error bars indicating cluster dispersions), for literature OCs from \citet{Reddy12, Reddy13, Reddy15} and \citet{Carretta07} adjusted to our abundance scale (squares), and Cepheids from \citet{And02a, And02b, And02c, And04} and \citet{Luck03} (triangles, Eu, Gd and Nd) and \citet{LuckLambert11} (triangles, Pr). The solid black line marks the best fit trend, and gray lines mark the 95\% confidence interval on the slope.}
\label{age_trends}
\end{figure*}

\indent In Figure \ref{age_trends}, we plot our [n-/Fe] ratios vs. cluster age (circles) with error bars marking the cluster dispersions (for clusters with more than one star) or the stellar dispersions (for clusters with only one star). Also plotted are literature OC measurements from \citet{Reddy12, Reddy13, Reddy15} and \citet{Carretta07} (squares) and Cepheid abundances from \citet{And02a, And02b, And02c, And04} and \citet{Luck03} for Eu, Gd, and Nd and \citet{LuckLambert11} for Pr (triangles). We have assumed a typical Cepheid age of 200 Myr, and the error bars on the Cepheid points represent the standard deviations of all Cepheid abundance measurements. The black line marks the best fit to our data, and the gray lines show the 95\% confidence interval on the slope.
\\
\indent If r-process elements trace the remnants of massive stars (type II supernovae or neutron stars, M$>$8M$_{\odot}$) and iron mainly traces lower mass supernovae (type Ia, M$<$8M$_{\odot}$), we might expect that through the history of the Galaxy early clusters will form with high levels of r-process enrichment from a few massive stars but progressively higher levels of iron as longer and longer-lived low-mass stars pollute the ISM. Thus, we may see a decrease in [r-/Fe] for young clusters. The selected literature sample does not have the same coverage of intermediate and old OCs as ours (the single old OC at 10 Gyr is NGC 6791) but it is consistent with our data set. 
\\
\indent Table \ref{tab:fit_params} shows the linear regression fit parameters for our abundance trends with age and Galactocentric radius. The [Eu/Fe] trend with age based on our OC data is on the edge of statistical significance (p = 0.039) with a slope of 0.024 $\pm$ 0.011 dex Gyr$^{-1}$. \citet{JF13} find a similar [Eu/Fe] trend with age of 0.023 dex Gyr$^{-1}$ but with a higher p = 0.074. We were able to reduce the dispersions in clusters for which we obtained additional spectra (N1817, N2141, N6939), and the addition of N6192 at 0.18 Gyr also makes the trend in the expanded data set more compelling. \citet{Yong12} find an age trend in their sample [Eu/Fe] plus literature data of -0.01 $\pm$ 0.01 dex Gyr$^{-1}$, but the age range of their clusters is $\sim$2 to 7 Gyr with a handful of literature clusters outside of that age range. Our trend with age is influenced by nine clusters younger than 2 Gyr which display significantly lower [Eu/Fe] as a group than the intermediate age clusters 2 to 7 Gyr. On average, we find the younger clusters have [Eu/Fe] $\sim$0.2 dex lower than the intermediate ages. Thus, the difference in trends could be a cluster selection effect. It is also possible that the observed element-to-iron ratios have a non-linear relationship with age, as the decrease in [Eu/Fe] for clusters younger than 2 Gyr suggests there might be a break there, but the scatter in our abundance measurements is too large to allow us to evaluate goodness of fit for different models.
\\
\indent We also recalculated the best fit to the [Eu/Fe] abundances with age using only stellar measurements of the 6645$\mathrm{\AA}$ line. As mentioned previously, most literature studies rely only on the 6645$\mathrm{\AA}$ line when measuring Eu abundances in metal-rich stars. We find that this makes $\sim$0.001 dex Gyr$^{-1}$ difference to the slope and the error on the slope. The y-intercept changes by 0.01 dex. We conclude that including other, more blended lines isn't distorting our determined Eu trend with age.
\\
\indent The Cepheid [Gd/Fe] measurements are similar to values we find for our young open clusters. These data cannot be well placed in the context of the enrichment of the Galaxy with time like the OC data, but it is heartening that they reproduce the same abundance range as our data for this relatively understudied element. Unfortunately, due to the limited number of clusters for which we could measure Gd, the trend is not statistically significant (p = 0.240) although it is similar in strength to Eu, with a slope of 0.026 $\pm$ 0.021. We must be cautious about interpreting such limited data, but the fact that the cluster abundances for Eu and Gd track each other is interesting.
\\
\indent Due to the wavelength and signal-to-noise restrictions of our data, only eight clusters have Dy measurements presented here. The [Dy/Fe] data show a stronger and more significant increase with age than Eu or Gd, with a slope of 0.054 $\pm$ 0.017 and p = 0.018, although it should be noted that the [Dy/H] abundances are all within 0.1 dex of solar. Once again, the sparse data requires careful interpretation, but the Dy and Gd abundances track each other very well, as we would expect of the r-process elements of similar weight. If the r-process does indeed have multiple sites, those elements and isotopes with similar neutron numbers should still have roughly the same yields.
\\
\indent Because 32\% of solar Mo is produced by the r-process (see Section \ref{subsec:Mo}), it is not considered a majority s-process element but would be expected to display abundance patterns somewhere between the pure r- and s-process elements. We find that [Mo/Fe] shows a statistically insignificant increase with decreasing cluster age, the best linear fit to the data having a slope consistent with zero (-0.012 $\pm$  0.012 dex Gyr$^{-1}$). The trend is also influenced by the oldest cluster, Be 17, which has abundances that fall slightly below the best fit line for all elements measurable in the KPNO data. We must not place too much weight on the effects of this single cluster; when it is excluded, there is no visual suggestion of a trend with age.
\\
\indent Pr and Nd are about half-and-half r- and s- process, so again we would expect their trends with age to fall between the pure r- and s- process elements similar to Mo. Both exhibit statistically insignificant trends (p = 0.321 and 0.537) and their slopes are consistent with zero within the errors (-0.011 $\pm$  0.011 and -0.008 $\pm$  0.012 dex Gyr$^{-1}$, respectively). Be 17 also falls below the trend line for these elements, and when it is removed these element-to-iron ratios appear to have no relationship with cluster age. The Nd literature OC and Cepheid data match the abundances of our youngest clusters well. The Pr Cepheid data falls a bit below the typical abundances for young clusters, but the dispersion among the individual Cepheids for both Pr and Nd is consistent with the dispersion among the OCs.
\\
\indent If we consider the \citet{JF13} s-process trends with age for Zr, La, and Ba (-0.011, -0.015, and -0.036 dex Gyr$^{-1}$ respectively), then it seems the mixed r- and s- process elements do fall between the trends on the pure r- and s- elements, although the uncertainties on the slopes of the mixed elements prevent any strong conclusions. We find the dispersion about the trend lines with age for the neutron-capture elements is $\sim$0.10-0.15 dex, which is larger than many of the cluster dispersions but similar to the size of the errors expected from atmospheric parameter uncertainties. It is also about the same size as the abundance change in the trend line over the 10 Gyr age range of our clusters, offering further evidence that the mixed elements have no trends with age. 
\\
\indent It is interesting to note that our Be 31 abundances (based on a single star) are the highest of any OC for the three elements that had lines in the limited wavelength region for this stellar spectrum. The Be 31 [Eu/Fe] abundance is only slightly higher than Be 22, but its [Mo/Fe] and [Nd/Fe] abundances are $\sim$0.2 dex higher than any other cluster in our sample. Because it is in the intermediate age range (2.6 Gyr) it does not have a large effect on element trends with age, but its location in the outer disk does strongly influence determined trends with Galactocentric radius which we discuss in the next section.

\floattable
\begin{deluxetable}{l r r r r r r r r}
\tabletypesize{\scriptsize}
\tablewidth{0pt}
\tablecolumns{9}
\tablecaption{Cluster Linear Regression Parameters  \label{tab:fit_params}}
\tablehead{\colhead{Variables} & \colhead{Slope} & \colhead{$\sigma_{slope}$} & \colhead{Intercept} & \colhead{$\sigma_{int.}$} & \colhead{R} & \colhead{P-value} & \colhead{N} & \colhead{$\sigma_{resid.}$}}
\startdata
{[}Mo/Fe] vs. Age & -0.012 & 0.012 & 0.150 & 0.051 & -0.222 & 0.310 & 23 & 0.141 \\
{[}Pr/Fe] vs. Age & -0.011 & 0.011 & 0.074 & 0.048 & -0.222 & 0.321 & 22 & 0.130 \\
{[}Nd/Fe] vs. Age & -0.008 & 0.012 & 0.132 & 0.053 & -0.136 & 0.537 & 23 & 0.145 \\
{[}Eu/Fe] vs. Age & 0.024 & 0.011 & -0.133 & 0.047 & 0.434 & 0.039 & 23 & 0.129 \\
{[}Gd/Fe] vs. Age & 0.026 & 0.021 & 0.050 & 0.084 & 0.437 & 0.240 & 9 & 0.122 \\
{[}Dy/Fe] vs. Age & 0.054 & 0.017 & -0.058 & 0.071 & 0.795 & 0.018 & 8 & 0.097 \\
{[}Fe/H] vs. R$_{\mathrm{GC}}$ & -0.047 & 0.007 & 0.439 & 0.078 & -0.831 & 0.000 & 23 & 0.074 \\
{[}Mo/Fe] vs. R$_{\mathrm{GC}}$ & 0.015 & 0.013 & -0.065 & 0.149 & 0.250 & 0.250 & 23 & 0.140 \\
{[}Pr/Fe] vs. R$_{\mathrm{GC}}$ & 0.039 & 0.011 & -0.393 & 0.122 & 0.624 & 0.002 & 22 & 0.104 \\
{[}Nd/Fe] vs. R$_{\mathrm{GC}}$ & 0.039 & 0.011 & -0.331 & 0.122 & 0.623 & 0.001 & 23 & 0.115 \\
{[}Eu/Fe] vs. R$_{\mathrm{GC}}$ & 0.039 & 0.010 & -0.492 & 0.116 & 0.648 & 0.001 & 23 & 0.109 \\
{[}Gd/Fe] vs. R$_{\mathrm{GC}}$ & 0.038 & 0.019 & -0.315 & 0.232 & 0.603 & 0.086 & 9 & 0.109 \\
{[}Dy/Fe] vs. R$_{\mathrm{GC}}$ & 0.050 & 0.020 & -0.457 & 0.241 & 0.716 & 0.046 & 8 & 0.111 \\
\enddata
\end{deluxetable}

\subsection{Galactocentric Radius Effects}  \label{subsec:radius}
\subsubsection{Iron}  \label{subsubsec:rad_iron}

\begin{figure*}
\epsscale{0.6}
\plotone{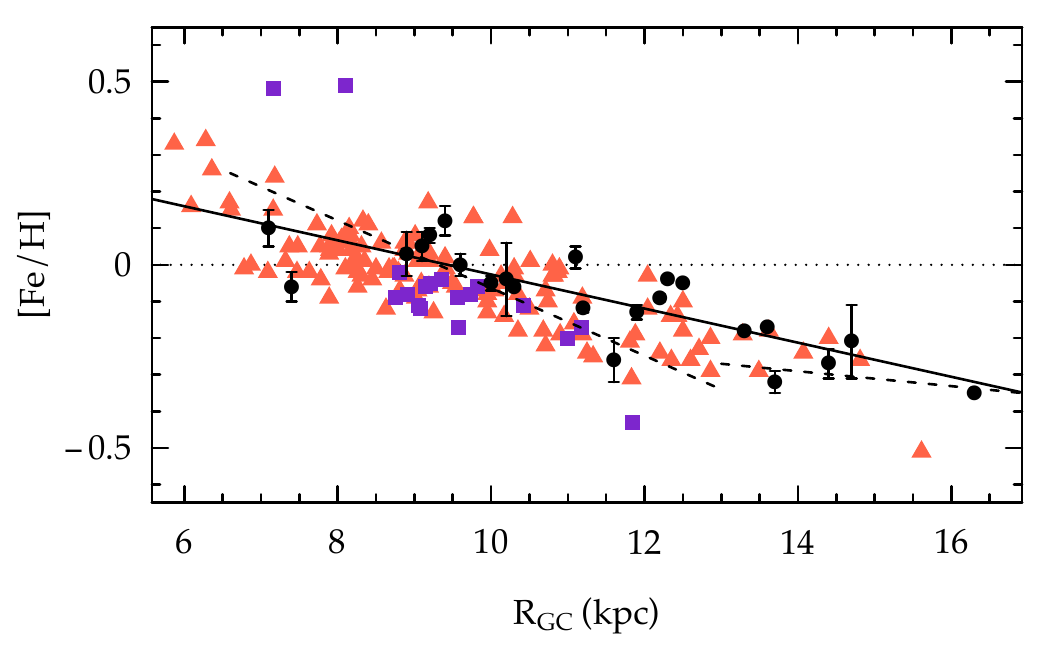}
\caption{Our cluster [Fe/H] abundances with Galactocentric radius; the solid line represents a linear fit to the data (see text) and the dashed lines are an iron gradient fit with a break at 13 kpc from \citet{Yong12}. Symbols are the same as Figure \ref{age_trends} except here triangles represent individual Cepheid measurements.}
\label{Fe_rgc}
\end{figure*}

\indent In Figure \ref{Fe_rgc} we plot [Fe/H] abundances for our sample clusters, selected literature OC data, and \citet{And02a, And02b, And02c, And04} and \citet{Luck03} Cepheid data against Galactocentric radius, using the same symbols as Figure \ref{age_trends}. Our clusters have R$_{\mathrm{GC}}$ $\sim$ 7 to 16 kpc, with the furthest cluster being Be 31. Several studies of the Galactic Fe gradient have noted a change in the gradient at $\sim$10 to 13 kpc \citep{Twarog97, Friel10, CP11, Yong12}. Our data do not extend far enough into the outer disk to assess whether our measured [Fe/H] distribution is indeed best fit by two linear regressions instead of one, so in Figure \ref{Fe_rgc} we mark a single linear regression to our OC data set as a solid line, and the [Fe/H] fit with R$_{\mathrm{GC}}$ with a break at 13 kpc from \citet{Yong12} as dashed lines. Our cluster sample also does not include any extremely high metallicity OCs, but \citet{Carretta07} have two inner disk clusters with [Fe/H] $>$ 0.4 dex, NGC 6253 and NGC 6791. These are old open clusters with ages of $\sim$3 and 10 Gyr and abundances based on 4 member stars each; NGC 6791 is the only open cluster in the sample of \citet{Salaris04} found to be older than Berkeley 17.
\\
\indent Supplementing our [Fe/H] data with Cepheid data over the same distance range, we do see a suggestion of a leveling out around 13 kpc. The Cepheid [Fe/H] abundances also decrease more sharply with R$_{\mathrm{GC}}$ than our OC abundances. The slope of the [Fe/H] gradient found for R$_{\mathrm{GC}}$ $<$ 13 kpc in \citet{Yong12} is -0.09 dex Gyr$^{-1}$; from our OC data alone, we find -0.047 $\pm$ 0.007 dex Gyr$^{-1}$, but the \citet{And02a, And02b, And02c, And04} and \citet{Luck03} Cepheid data gives $\sim$ -0.06 dex Gyr$^{-1}$. Our clusters fall roughly in line with the Cepheid [Fe/H] at different radii, so the difference in slopes may be due to the relative sparseness of inner disk OCs in our sample.

\subsubsection{r-process elements}  \label{subsubsec:rad_r}

\begin{figure*}
\epsscale{0.6}
\plotone{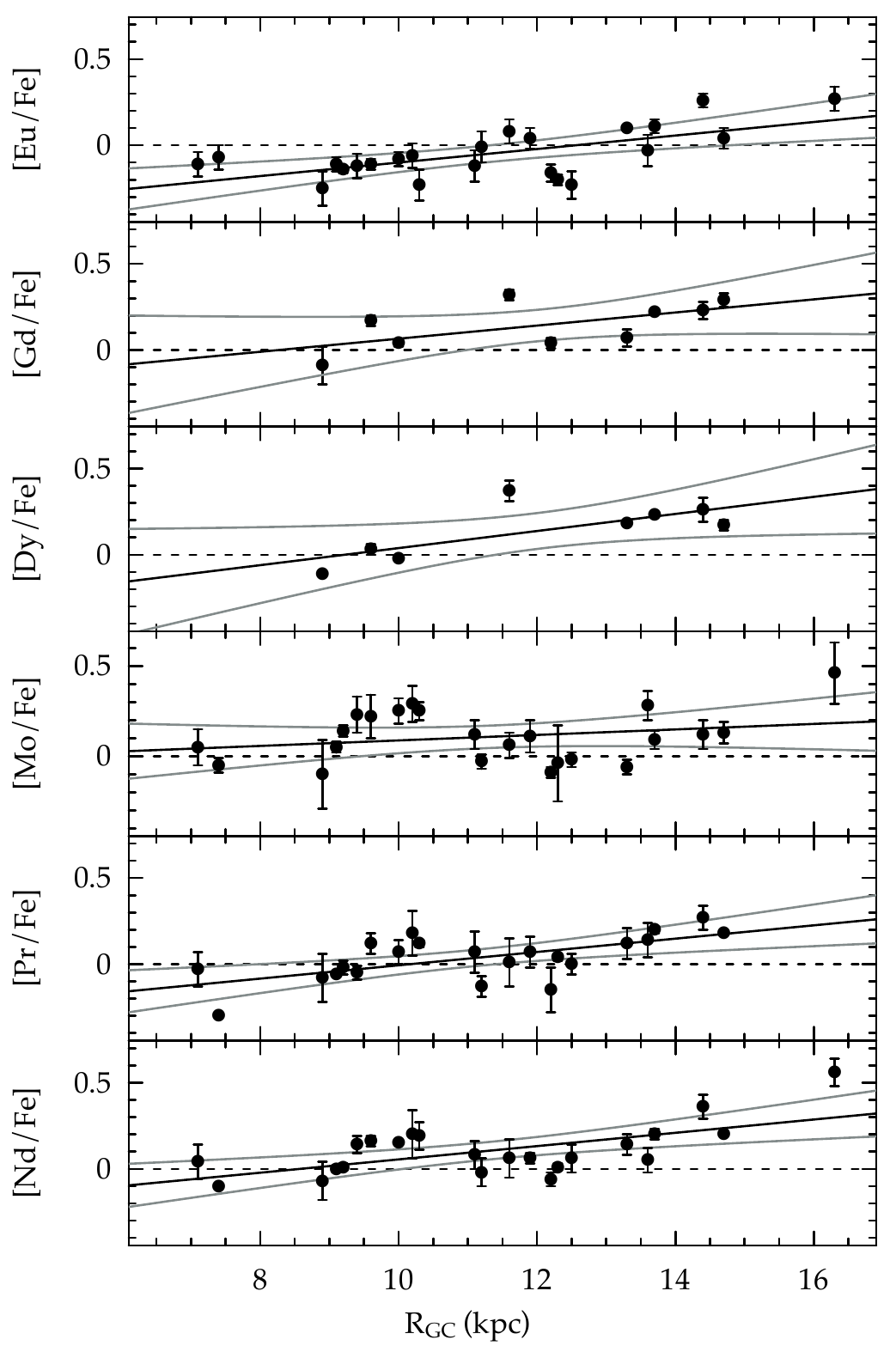}
\caption{Abundance trends with Galactocentric radius; symbols and lines the same as in Fig. \ref{age_trends}.}
\label{rgc_fits}
\end{figure*}

\begin{figure*}
\epsscale{0.6}
\plotone{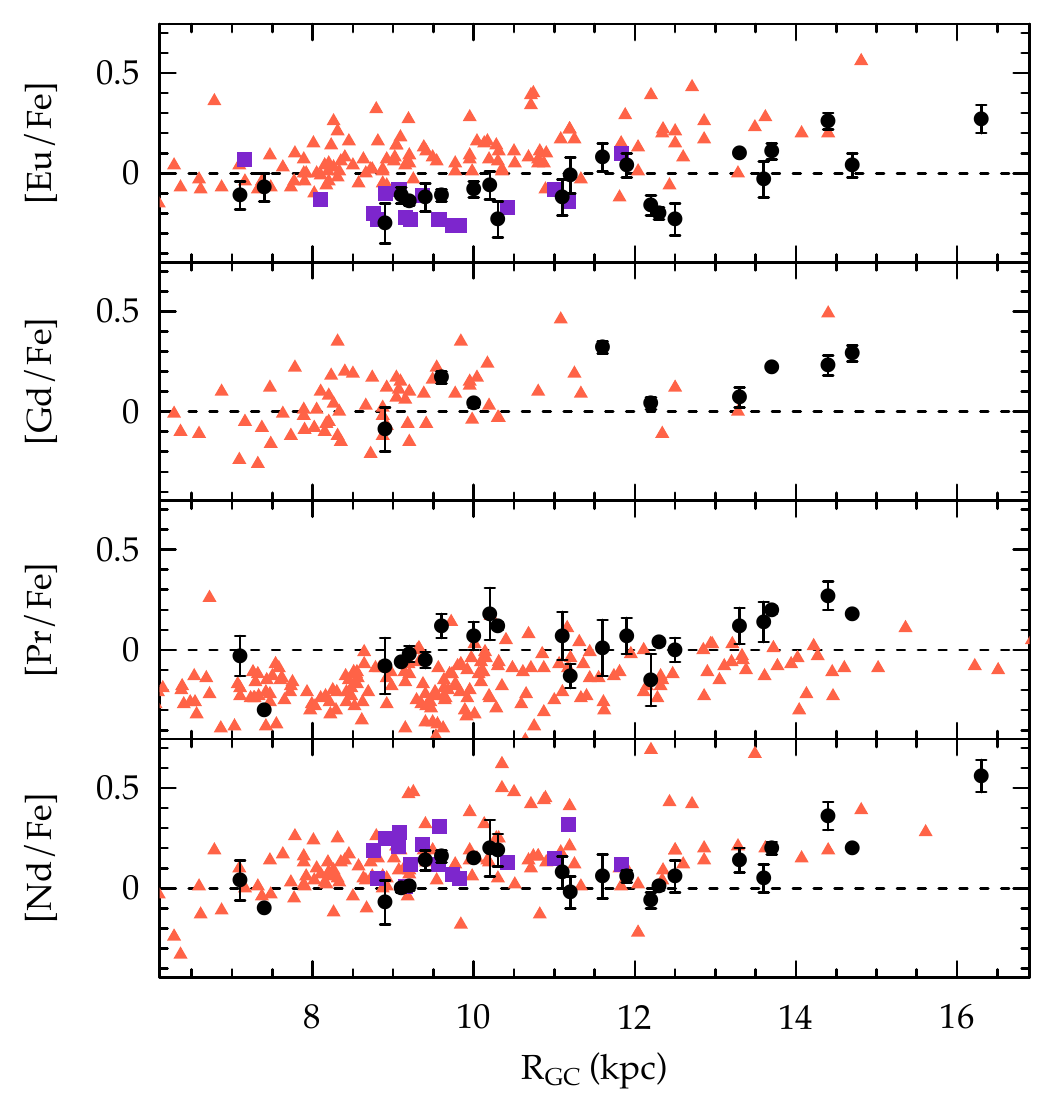}
\caption{Abundances and literature data with Galactocentric radius; symbols and lines the same as in Fig. \ref{age_trends}, except that triangles now represent individual Cepheid measurements.}
\label{rgc_lit}
\end{figure*}

\indent In Figure \ref{rgc_fits}, we plot our [n-capture/Fe] ratios vs. Galactocentric radius (circles) with error bars marking the cluster dispersions (for clusters with more than one star) or the stellar dispersions (for clusters with only one star), the black lines marking the best fit to our data, and the gray lines showing the 95\% confidence interval on the slopes.
\\
\indent We find a [Eu/Fe] with R$_{\mathrm{GC}}$ trend of 0.039 $\pm$ 0.010 dex kpc$^{-1}$ and p-value of 0.001, which seems to be driven by clusters beyond 13 kpc. \citet{JF13} find a trend with R$_{\mathrm{GC}}$ of 0.047 dex kpc$^{-1}$ with a p-value of 0.017. \citet{Yong12} fit two regressions to [Eu/Fe] with R$_{\mathrm{GC}}$ for the apparent break in [Fe/H] at 13 kpc, but since our clusters do not extend as far into the outer disk we do not calculate the separate regressions; the slope of their regression for R$_{\mathrm{GC}}$ $<$ 13 kpc is 0.07 $\pm$ 0.01 dex kpc$^{-1}$, and for R$_{\mathrm{GC}}$ $>$ 13 kpc is 0.01 $\pm$ 0.00 dex kpc$^{-1}$. Their sample covers mainly intermediate-distance clusters with 11 $<$ R$_{\mathrm{GC}}$ $<$ 16 kpc, with one inner and outer cluster from their data set and four inner disk clusters drawn from the literature. 
\\
\indent Attempting to correct for all systematic differences between their sample and the literature was not the goal of the work, so there may be some remaining differences causing a disparity between their intermediate R$_{\mathrm{GC}}$ clusters and inner disk clusters from the literature. As discussed previously, due to differences in atomic data and line measurement techniques, we have seen variations in [Eu/Fe] measurements up to 0.3 dex between sources. If we consider a break in the [Eu/Fe] gradient at 13 kpc, we find an insignificant inner gradient of 0.003 $\pm$ 0.014 dex kpc$^{-1}$ and a strong outer gradient for the handful of clusters we have of 0.069 $\pm$ 0.042 dex kpc$^{-1}$. However, considering only the clusters with [Eu/Fe] measurements by \citet{Yong05, Yong12} over the Galactocentric radius range of our data, we find a best fit slope of 0.031 $\pm$ 0.023 dex kpc$^{-1}$, which is consistent with our single regression slope.
\\
\indent We have again calculated the [Eu/Fe] trend with R$_{\mathrm{GC}}$ based only on the 6645$\mathrm{\AA}$ line. We find no significant difference from the trend derived from all four lines; [Eu/Fe]$_{6645}$ = 0.035 $\pm$ 0.010 x R$_{\mathrm{GC}}$ - 0.430 $\pm$ 0.114.
\\
\indent Our sample [Gd/Fe] trend with R$_{\mathrm{GC}}$ matches the [Eu/Fe] trend, though with a lower statistical significance, with a slope of 0.038 $\pm$ 0.019 dex Gyr$^{-1}$ and p-value of 0.086. Our [Dy/Fe] OC trend is stronger than Gd or Eu, with a slope of 0.050 $\pm$ 0.020 dex Gyr$^{-1}$, p = 0.046.
\\
\indent Figure \ref{rgc_lit} plots our data over literature data with R$_{\mathrm{GC}}$; sources are represented with the same symbols as Figure \ref{age_trends}, but with triangles now representing individual Cepheid values. The literature OC abundances for [Eu/Fe] with R$_{\mathrm{GC}}$ match our determined abundances well. 
\\
\indent The Cepheid [Eu/Fe] data trend with R$_{\mathrm{GC}}$ has a slope of 0.035 $\pm$ 0.005 dex kpc$^{-1}$ which is similar to our calculated value although there does appear to be some systematic offset between the two data sets. Because the Cepheid abundances were measured using only equivalent widths, this may create some systematic differences for the elements we have synthesized using hyperfine or isotopic structure.
\\
\indent The Cepheid [Gd/Fe] measurements do not cover the full Galactocentric radius range of our cluster sample, but they are similar to values we find for our open clusters over the range of the Cepheid data ($\sim$7-11 kpc). The Cepheid [Gd/Fe] trend with R$_{\mathrm{GC}}$ matches our OC trend well with a slope of 0.041 $\pm$ 0.010 dex Gyr$^{-1}$ and p-value of $\sim$0.0001.

\subsubsection{Mixed r- and s- elements}  \label{subsubsec:rad_mixed}

\begin{figure*}
\epsscale{0.8}
\plotone{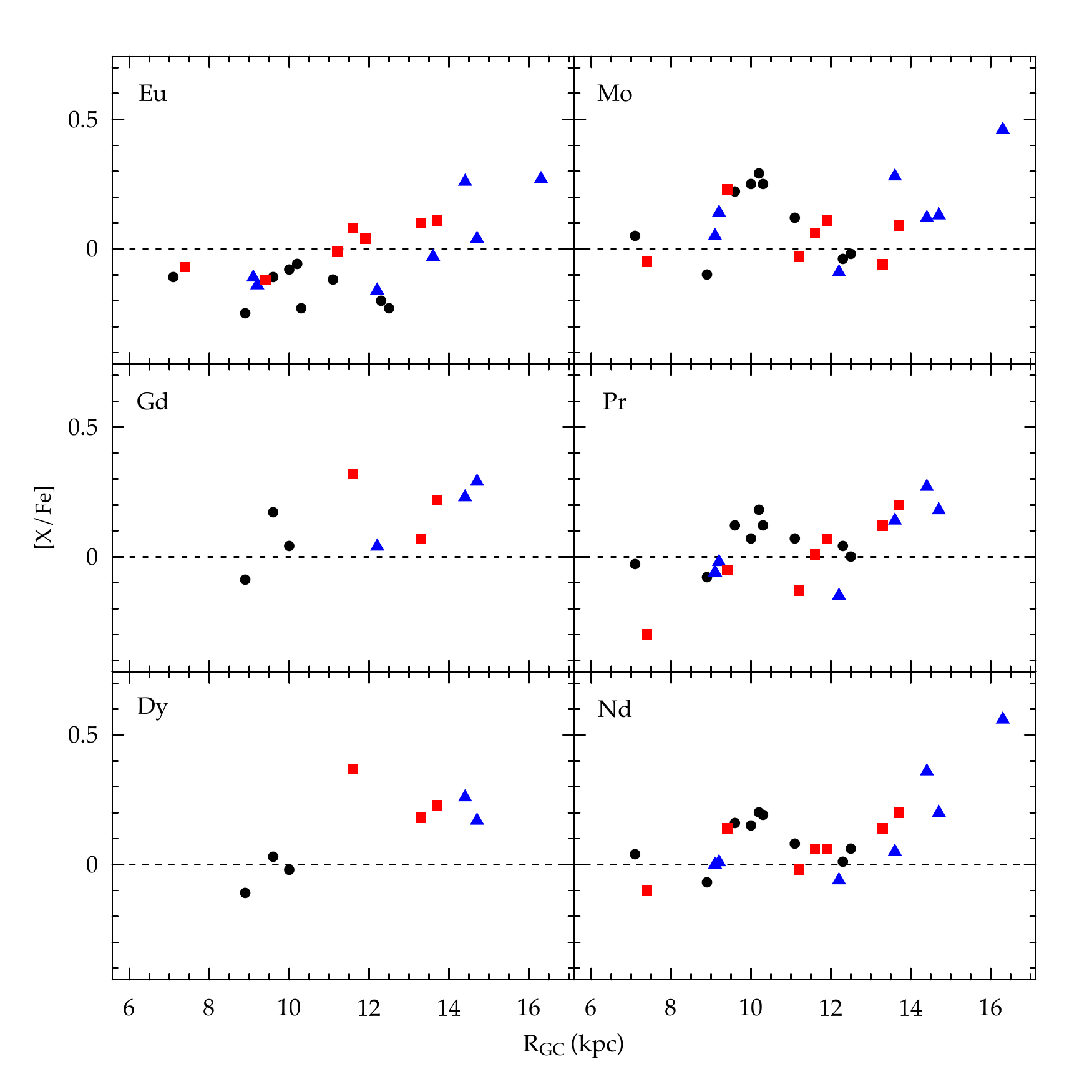}
\caption{Same data as Fig. \ref{rgc_fits} but without error bars and separated by ages. Circles are clusters younger than 2 Gyr, triangles are 2 $<$ age $<$ 5 Gyr, and squares are older than 5 Gyr.}
\label{rgc_age}
\end{figure*}

\indent [Mo/Fe] has a statistically insignificant trend with Galactocentric radius, with a slope of 0.014 $\pm$ 0.013 dex kpc$^{-1}$ (p = 0.306). There is little comparison data available for this element and the limitations on the determination of the trend with R$_{\mathrm{GC}}$ means there is little we can say about its behavior. 
\\
\indent [Pr/Fe] and [Nd/Fe] display much more significant trends with Galactocentric radius of 0.037 $\pm$ 0.011 dex kpc$^{-1}$ with p-values of 0.002 and 0.003 respectively. It is reassuring that trends for these elements are similar; Pr has no measurement for Be 31, but the fact that the results are still similar means that this cluster may not have a serious impact on the overall results. Praseodymium also has some Cepheid data available in the literature from \citet{LuckLambert11}, which displays a moderate R$_{\mathrm{GC}}$ gradient similar to our data (0.04 dex kpc$^{-1}$) but does not match our abundance scale. Again, the discrepancy in the scale of the data is probably due to our use of hyperfine structure in synthesis. The literature OC and Cepheid data for Nd both match our data well; [Nd/Fe] shows a suggestion of an upturn at 13 kpc that is mirrored in the Cepheid data.
\\
\indent The cluster averages of the mixed r- and s- elements, Mo, Pr, and Nd, display significant scatter around the linear regression that cannot be explained by the dispersions of individual clusters. It is true that the standard deviations of individual stellar abundance measurements are an underestimation of the true cluster errors, particularly for clusters with two or three stars (error bars on clusters for which we have only one star are the stellar abundance dispersions), but there are still some clusters that fall far outside the 95\% confidence interval on the slope (gray lines on Figs. \ref{age_trends} and \ref{rgc_fits}). Though the scatter about the trends with age mostly seems random, there is some visible structure in the relationship of cluster abundances with galactocentric radius.
\\
\indent For Figure \ref{rgc_fits}, we have assumed a simple linear relationship between [X/Fe] and R$_{\mathrm{GC}}$, but this probably does not accurately reflect the process of Galactic enrichment. Both possible r-process events (NS-NS mergers and type II supernovae) occur rarely and even for young stars the local ISM may not have been enriched by many such events; \citet{Kalogera04} estimate a current Galactic neutron star merger rate of 180 Myr$^{-1}$. \citet{Diehl06} estimate a Galactic CCSN rate of 19000 Myr$^{-1}$, though a recent chemical evolution model by \citet{Wehmeyer15} estimates that only 0.1\% of these are magnetorotationally driven `jet' supernovae which would be capable of producing and ejecting significant amounts of r-process material. Thus, we may expect local inhomogeneities in the ISM. 
\\
\indent The mixed elements in particular seem to have some structure with Galactocentric radius: there is a slight enhancement of about 0.2 dex from the linear regression around 10 kpc from the Galactic center, a decrease of about the same amount at 12 kpc, and then another increase out to large radii. The increase at around 15 kpc is based on a single cluster for which we only have one star (Be 31), which has no Pr abundance determined because of wavelength restrictions (see Section \ref{subsec:Pr}). The increase and decrease at 10 and 12 kpc, however, involve several clusters varying from a linear relationship. The available Cepheid data show a dispersion of $\sim$0.15 dex within bins of 1 kpc in Galactocentric radius; if we use this as a marker of intrinsic abundance scatter in the ISM at any given radius, Be 31 is the only cluster that really stands out. The other clusters have residuals of up to 0.2 dex, but these variations might be explained by the local abundance variations in the ISM or errors due to atmospheric parameters and continuum setting. Further data for s-process elements may help evaluate the significance of these nonlinearities.
\\
\indent In Figure \ref{rgc_age}, we show [X/Fe] vs. Galactocentric radius for all elements, with clusters color-coded by age. For elements with the full set of cluster abundances, there seems to be no trend with Galactocentric radius in the youngest clusters (age $<$ 2 Gyr, circles), though the intermediate age (2 - 5 Gyr, triangles) and old ($>$ 5 Gyr, squares) cluster abundances increase with Galactocentric radius for [Eu/Fe], [Pr/Fe], and [Nd/Fe]. More specifically, when fitting a linear regression to each age group individually, the youngest clusters have a slope consistent with zero for all elements, the intermediate age clusters have positive slopes from 0.04 (Mo) to 0.06 (Nd) $\pm$ 0.02 dex kpc$^{-1}$, and the oldest clusters have positive slopes for Eu (0.035 $\pm$ 0.009), Pr (0.070 $\pm$ 0.013), and Nd (0.034 $\pm$ 0.015). 
\\
\indent Because the youngest clusters mostly cover the inner disk (R$_{\mathrm{GC}}$ $\le$ 12 kpc) it is difficult to determine whether the differences in gradients for the age groups are due to mixing (or some other long-timescale effect), or whether the outer disk has a distinct star formation history from the inner disk (i.e. whether the younger clusters would show increased neutron-capture abundances if data were available at large radii). \citet{Bird12} use simulations of perturbed and unperturbed Galactic disks to estimate a typical radial migration of $\sim$1 kpc for disk stars, with 80\% moving less than 2 kpc in R$_{\mathrm{GC}}$. They note that the perturbed disks experience the largest stellar migrations preferentially at large radii, where the disk becomes unstable; a significant fraction of outer disk stars with initial radii $>$ 15 kpc ($\sim$40\%) migrate more than 3 kpc. It is possible that mixing due to radial migration would require long timescales on the order of Gyr to reach equilibrium.
\\
\indent \citet{Wu09} calculate orbits for a large sample of OCs, twelve of which are also in our sample \citep[we do not include Be 31 orbital calculations because of issues with its available proper motion data; see][]{VandePutte10}. A few of these are highly eccentric; in particular, they find that Be 21 has an eccentricity of 0.47, and calculate that it varies over 8 kpc in its orbit from apogalacticon to perigalacticon. The median difference between R$_{\mathrm{apo}}$ and R$_{\mathrm{per}}$ for the dozen clusters in common is $\sim$3 kpc. This variation may also contribute significantly to the apparent `mixing' moderating the radial gradients presented here.
\\
\indent There is not enough data for Gd and Dy to examine Galactocentric radius trends in the three age groups, but it is worth noting that the youngest clusters, which all fall in the 8-10 kpc range in this cluster subset, appear to be driving the overall relationship with R$_{\mathrm{GC}}$. 

\subsection{Element-to-Element Ratios}  \label{sec:ratios}

\begin{figure*}
\epsscale{0.6}
\plotone{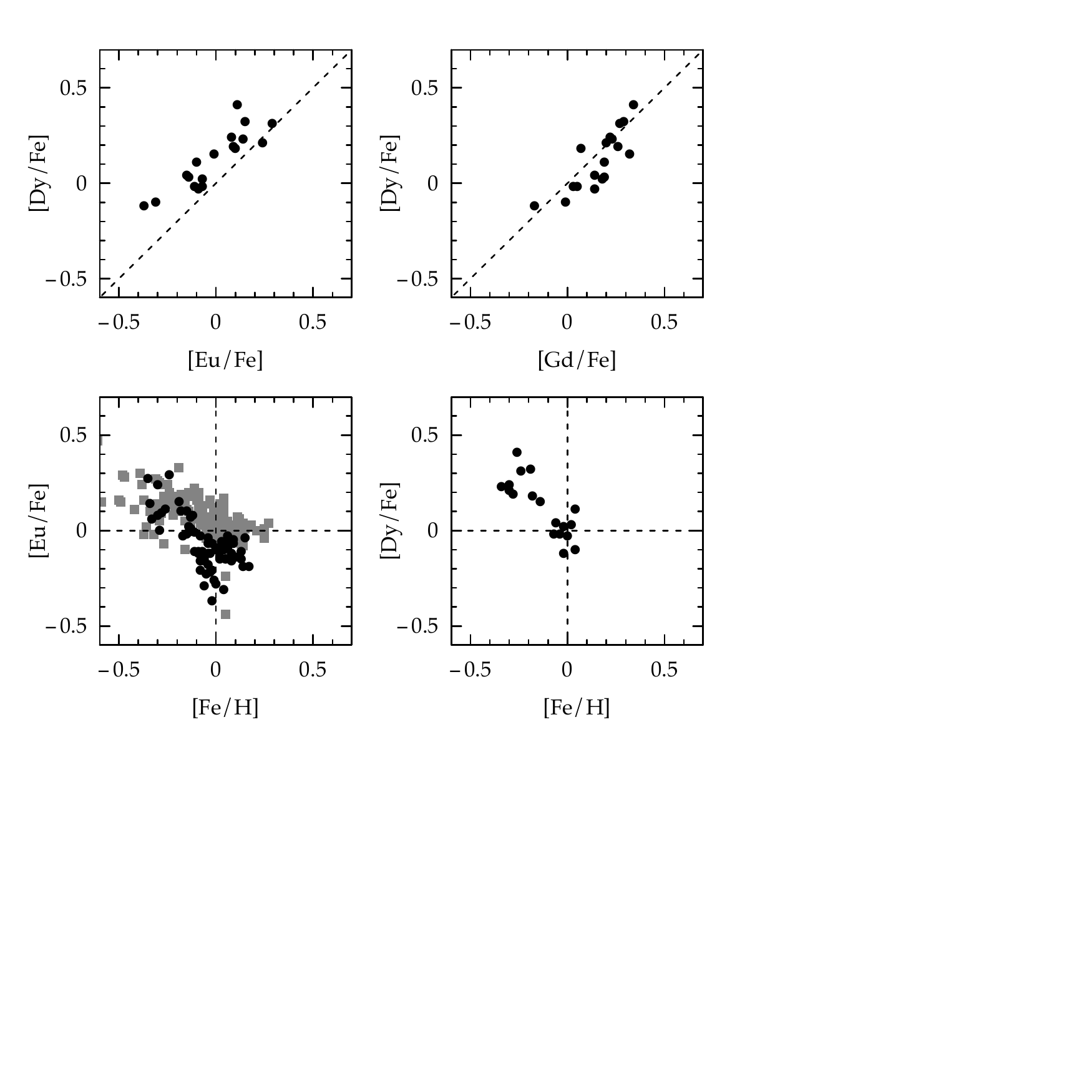}
\caption{Stellar abundance vs. abundance plots for the r-process elements (Eu, 98\% r; Gd, 82\% r; Dy, 88\% r) and Fe. Circles are our abundances, and squares are disk giant abundances from \citet{Mishenina07}.}
\label{r_ratios}
\end{figure*}

\begin{figure*}
\epsscale{0.9}
\plotone{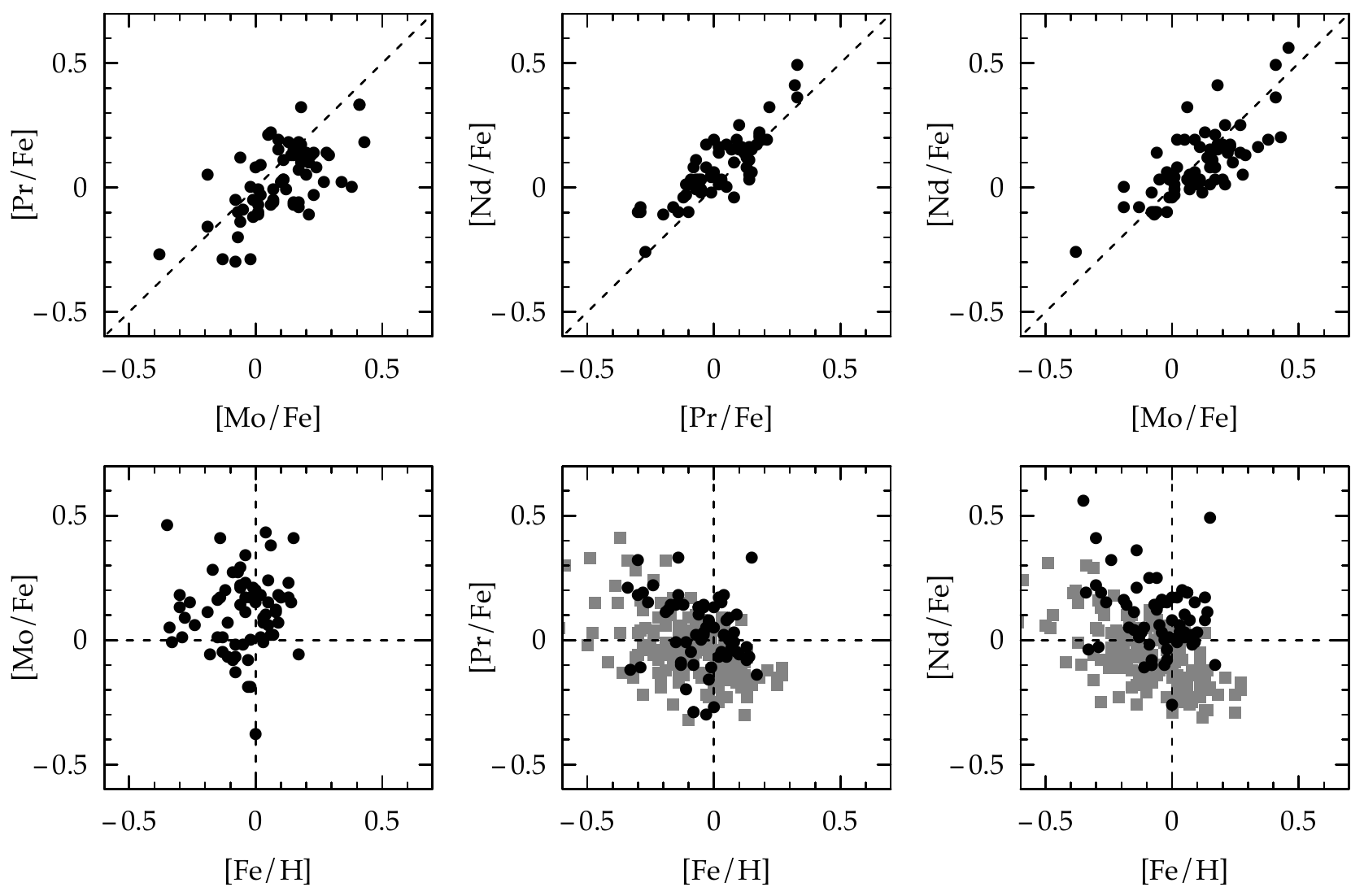}
\caption{Abundance vs. abundance plots for the mixed elements (Mo, 32\% r; Pr, 51\% r; Nd, 42\% r), and Fe. Symbols are the same as in Fig. \ref{r_ratios}.}
\label{mixed_ratios}
\end{figure*}

\begin{figure*}
\epsscale{0.9}
\plotone{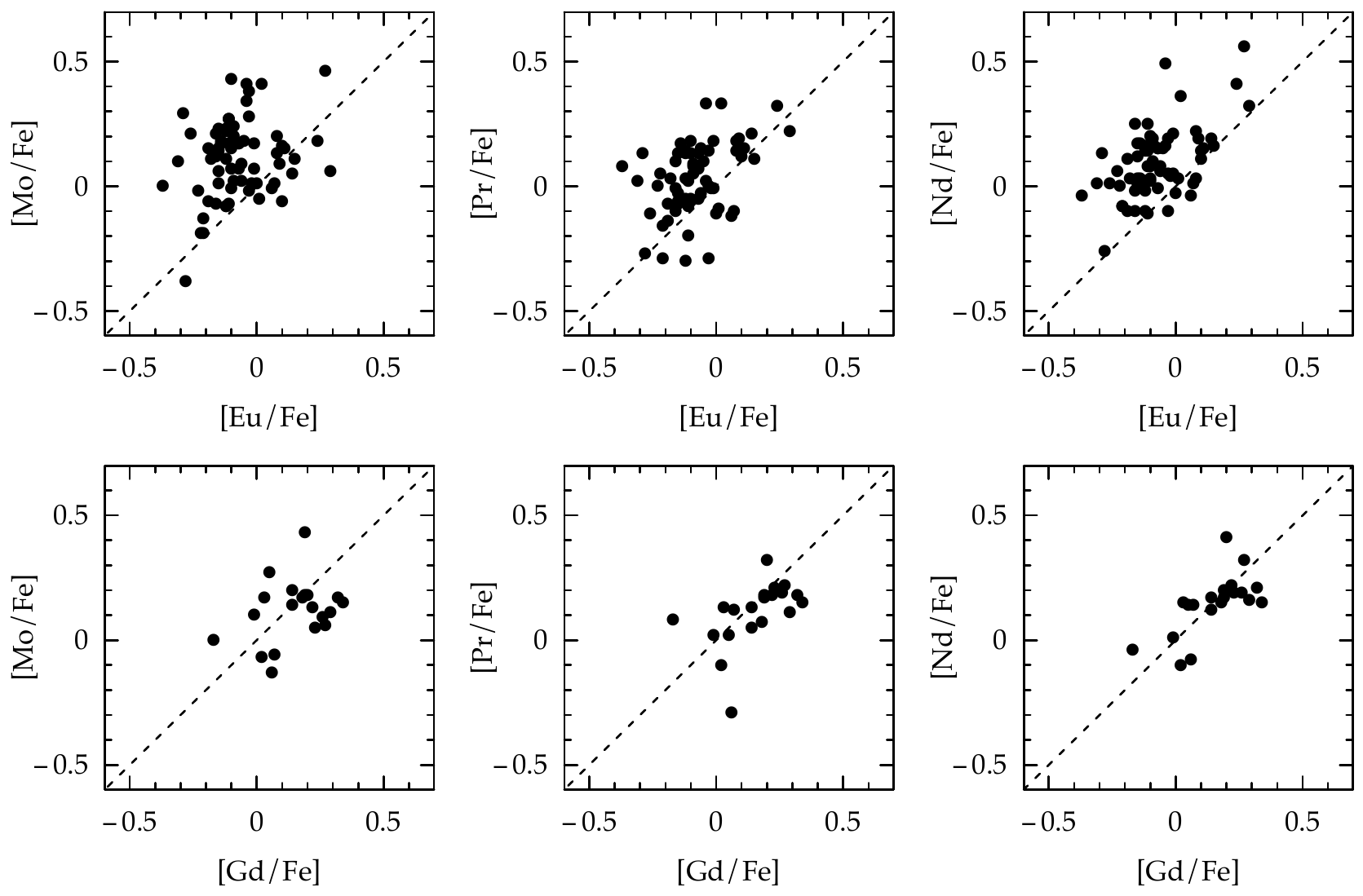}
\caption{Abundance vs. abundance plots for r-process and mixed elements (Eu, 98\% r; Gd, 82\% r; Mo, 32\% r; Pr, 51\% r; Nd, 42\% r).}
\label{r_mixed_ratios}
\end{figure*}

\indent One check on the quality of our abundance measurements is the degree to which cluster abundances for elements in similar groups scale with each other. Since our group of r-process elements were presumably formed via the same nucleosynthetic mechanism, we would expect that stars showing an enhancement in one r-process element would also be enhanced in other r-process elements, and the same for mixed elements with approximately the same level of r- and s- contributions. By the same reasoning, if systematic biases are not responsible for variations in abundance measurements from star to star, we would expect to see no correlation between r-process and s-process element abundances which are produced in different astrophysical sites.
\\
\indent Figure \ref{r_ratios} shows stellar r-process element abundances plotted against each other and [Fe/H] (dotted line is a 1:1 correspondence for r- vs r- plots and indicates solar ratios for r- vs. Fe plots). [Eu/Fe] points are plotted over disk field giant data from \citet{Mishenina07}. Though [Dy/Fe] and [Gd/Fe] track each other well, there appears to be an offset between these two and our Eu (top left and right panels) of $\sim$0.15 - 0.20 dex. However, in Figs. \ref{age_trends} and \ref{rgc_lit} we see that the Cepheid Eu and Gd abundances have similar abundances relative to solar. If these r-process elements are formed in the same environment/location, we have no reason to expect that they would give different abundances relative to solar; we may have scaling issues due to our choice of atomic data, since the 6645$\mathrm{\AA}$ line which most literature measurements are based on has significant hyperfine and isotopic structure, and different linelists based on many features may not give results that can be scaled simply to solar values, particularly for giant stars.
\\
\indent The \citet{Mishenina07} study of 171 cool disk giants gives [Eu/Fe] abundances systematically 0.1 to 0.2 dex higher than our values for the same [Fe/H]. Model calculations of thin disk [Eu/Fe] from \citet{Serminato09} also agree well with the \citet{Mishenina07} data for all but perhaps the highest [Fe/H] stars. [Eu/Fe] models from \citet{Cescutti06} predict a [Eu/Fe] of about +0.2 at [Fe/H] of -0.4 dex, $\sim$ +0.05 at solar [Fe/H], and -0.15 at [Fe/H] of +0.2 dex. In particular, our stellar [Eu/Fe] abundances at solar [Fe/H] that fall below about -0.25 dex are not matched in the literature. Some of these stellar [Eu/Fe] abundances are low compared to other members of the cluster; NGC 2158, NGC 2194, and NGC 6939 are the only clusters to have an average [Eu/Fe] below -0.2 dex, and their internal dispersions are large enough to be consistent with the lower edge of the disk star abundances.
\\
\indent Interestingly, \citet{Wehmeyer15} use a chemical model with non-instantaneous recycling and mixing to locate the r-process site by modeling the scatter in [Eu/Fe] with time for the history of the Galaxy. The modeling is focused on the early history of the Galaxy, as the authors note that a mixture of NS-NS mergers and jet supernovae ejecta would be required to reproduce the observed scatter in [Eu/Fe] at early times. However, they also predict a scatter in [Eu/Fe] at solar metallicities (-0.3 $<$ [Fe/H] $<$ 0.2 dex) of $\sim$ 0.4 to 0.5 dex. Some of the assumptions used to model the local mixing may not hold at later times in the evolution of the Galaxy, but the close match between the predictions and our observed scatter is intriguing.
\\
\indent Figure \ref{mixed_ratios} shows mixed element abundances plotted against each other and [Fe/H]. Nd and Pr correlate strongly with each other, but not as strongly with Mo. None of the three elements correlate strongly with [Fe/H]. The correlation of Pr and Nd (51 and 42\% r-process, respectively) is encouraging; the weaker correlation with Mo may be due to its smaller r-process contribution (32\%) or due to the smaller number of lines measured (3 vs. 4 and 10) and relatively large abundance errors (Mo has errors due to temperature uncertainties on the order of 0.15 dex, and large continuum errors due to the weak lines). The abundance errors due to the continuum uncertainty, which are as high as 0.3 dex for some stars, may be systematic and could explain the full extent of the deviation from a 1:1 relationship with Pr and Nd. However, based on our error estimates (Table \ref{tab:errors}) systematic effects could not explain the abundance ranges of $\sim$0.8 dex we see in the full sample.
\\
\indent \citet{Mishenina07} also give [Nd/Fe] and [Pr/Fe] abundances for giant stars. The [Pr/Fe] abundances range from +0.4 dex at an [Fe/H] of -0.4 to -0.3 dex at an [Fe/H] of +0.2, in good agreement with our stellar [Pr/Fe] abundances (the star at [Fe/H] of +0.15 and [Pr/Fe] of +0.33 is anomalous N6939 star 190, which is also strongly enhanced in Nd). However, the [Nd/Fe] values for giants fall from +0.3 dex at an [Fe/H] of -0.4 to -0.3 dex at an [Fe/H] of +0.2, again showing a systematic offset with respect to our data, although the range of the data in this [Fe/H] region is about 0.6 dex for both sets of abundances. It is worth noting that we only measured three of the same Nd lines as \citet{Mishenina07}, and the log(gf) values we used for these lines are systematically lower than theirs by about 0.1 dex. Their measured solar Nd abundance of 1.50 is very similar to ours, but differences in scaling that might not affect solar measurements could still be significant for giants.
\\
\indent Figure \ref{r_mixed_ratios} shows r-process and mixed element abundances plotted against each other. None of the mixed r- and s-process elements appear to correlate with Eu. Correlations of Pr and Nd with Gd are tighter around the 1:1 line, perhaps due to the subsection of clusters represented in the Gd abundances. Again, it is possible that these correlations could be due to systematic errors, but the fact that the elements with the most similar mix of r- and s- contributions are the most tightly correlated is suggestive.

\section{Summary and Conclusions}  \label{sec:summary}

\indent There is still debate over the details of heavy element nucleosynthesis, i.e. the neutron-capture process(es). The site of the r-process has not been conclusively identified, and the yields and relative importance of the known s-process sites are currently being revised due to recent measurements of highly supersolar Ba abundances in young open clusters \citep[e.g.,][]{DOrazi09, Maiorca12}. Old, metal-poor stars and some disk populations have abundance measurements for r-process elements in the literature, but large sets of internally consistent measurements for open clusters are rare. Because r-process elements are thought to be produced in the environments or remnants of massive stars (M $>$ 8M$_\odot$) and s-process elements mainly in low-mass stars (M $<$ 8M$_\odot$), the range and accuracy of determined ages of open clusters make them a valuable tool to study the time scales of production for both sets of elements.
\\
\indent We present here six neutron-capture elements (three primarily r-process, Eu, Gd, and Dy; and three mixed r- and s-process, Mo, Pr, and Nd) for a sample of 68 stars in 23 open clusters. The data are a mixture of spectra previously analyzed by us, obtained from others, and data we have newly gathered. We have used an automated synthesis fitting program of our own devising to reduce the time necessary to measure features with blends and internal structure, and to keep our abundances as homogenous as possible. We are able to measure at least three lines for every element presented, with individual stellar dispersions generally $<$ 0.10 dex.
\\
\indent We analyze trends for our sample in [X/Fe] vs. cluster age and Galactocentric distance, and find the following:
\begin{itemize}
\item Eu, Gd, and Dy have trends with increasing age of 0.024, 0.026, and 0.054 dex Gyr$^{-1}$, respectively, though the trend in Gd is not statistically significant.
\item Mo, Pr, and Nd have age trends consistent with zero, which places them in between positive age trends of pure r-process and negative age trends previously found for Ba.
\item All elements except Mo and Gd have significant linear trends with Galactocentric radius, with linear regression slopes $\sim$ +0.04 dex kpc$^{-1}$.
\item The relationship of mixed elements (Mo, Pr, Nd) with Galactocentric radius may not be linear, as these appear to be enhanced around 10 kpc and drop around 12 kpc. The youngest OCs (age $<$ 2 Gyr) show no relationship with R$_{\mathrm{GC}}$, while older clusters show a strong positive correlation for Pr and Nd.
\item For all elements, the scatter at different ages and Galactocentric radii is larger than our cluster dispersions.

\end{itemize}

The behavior of clusters at large Galactocentric radii (particularly Be 31) presents a puzzle. An enhancement beyond 13 kpc is seen for most elements, though this radius range is sparsely populated in our data. Unfortunately, these clusters are some of the most difficult to observe. Additional clusters at this distance (and additional stars for Be 31) would help distinguish between effects in the different age ranges of clusters (and possible mechanisms for this difference) and overall trends with R$_{\mathrm{GC}}$. The possible abundance enhancement at 10 kpc may also be an age effect, as it appears to be populated by the youngest clusters, and may even be a sampling effect as the enhancement is not much larger than typical R$_{\mathrm{GC}}$ dispersions observed in Cepheids. However, since it is seen mainly in the mixed r- and s- elements, pure s-process abundance measurements may elucidate the issue. We will explore s-process Galactic enrichment in an upcoming paper focusing on a set of majority s-process elements.

\acknowledgments
J. C. O. gratefully acknowledges support by the Indiana Space Grant Consortium through a Graduate Fellowship. We would like to thank Michael Briley and Laura Magrini for generously allowing us use of McDonald and VLT data, and also APO 3.5m telescope operator Jack Dembicky for help with observing sessions and instrument training.

\bibliography{rprocess_refs}

\end{document}